\documentclass[ALICE,manyauthors]{cernphprep}

\usepackage{graphicx}  
\usepackage{dcolumn}   
\usepackage{bm}        
\usepackage{amssymb}   
\usepackage{amsfonts}
\usepackage{graphics}
\usepackage{epsfig}
\usepackage[normalem]{ulem}
\usepackage[usenames]{color}
\usepackage{multirow}
\usepackage{hyperref}
\usepackage{color}  
\usepackage{lipsum}

\def\pt{\mbox{$p_{\rm T}$}}
\def\kt{\mbox{$k_{\rm T}$}}
\def\qinv{\mbox{$q$}}
\def\snn{\mbox{$\sqrt{s_{_{NN}}}$} }

\makeatletter

\newcommand{\Rmnum}[1]{\expandafter\@slowromancap\romannumeral #1@}
\makeatother

\begin{document}          

\begin{titlepage}
\PHnumber{2013-201}                 
\PHdate{October 28, 2013}

\title{Two- and Three-Pion Quantum Statistics Correlations 
in Pb-Pb Collisions at \snn= 2.76 TeV at the CERN Large Hadron Collider}
\ShortTitle{Two and Three-Pion Quantum Statistics Correlations}
\Collaboration{ALICE Collaboration%
         \thanks{See Appendix~\ref{app:collab} for the list of collaboration
                      members}}
\ShortAuthor{ALICE Collaboration}

\begin{abstract}
Correlations induced by quantum statistics are sensitive to the spatio-temporal extent as well as dynamics of particle emitting sources in heavy-ion collisions.  
In addition, such correlations can be used to search for the presence of a coherent component of pion production.  
Two- and three-pion correlations of 
same and mixed-charge are measured at low relative momentum to estimate the coherent fraction of charged pions in Pb-Pb collisions at \snn= 2.76 TeV at the CERN Large Hadron Collider with ALICE.  
The genuine three-pion quantum statistics correlation is found to be suppressed relative to the two-pion correlation based on the assumption of fully chaotic pion emission.  
The suppression is observed to decrease with triplet momentum.  
The observed suppression at low triplet momentum may correspond to a coherent fraction in charged pion emission of $23\%\pm8\%$.

\end{abstract}
\end{titlepage}
\setcounter{page}{2}

\section{Introduction}

The techniques of intensity interferometry are often used to extract information of the space-time structure of particle-emitting sources 
\cite{QSbasics}.  For identical boson correlations, quantum statistics (QS) or Bose-Einstein correlations contribute significantly at low relative momentum.  
The strength of QS correlations is known to depend on the degree of chaoticity of particle-emitting sources \cite{WeinerCoherence,Gyulassy}.
Identical boson QS correlations reach their maximum value for fully chaotic sources (no coherence) and their minimum value for fully coherent sources.  The possibility of coherent pion 
production in high-energy heavy-ion collisions has been considered several times before.  In particular, it was proposed that the interior of the high-energy hadron collisions 
might form a Bose-Einstein condensate \cite{FowlerStelteWeiner} with an anomalous chiral order parameter (DCC) \cite{BjorkenAlaska}.  
Such a condensate produced in the interior may survive until some time after the relatively hot and chaotic expanding shell decouples and hadronizes.
The pion radiation from a condensate is expected to be coherent and thus suppresses Bose-Einstein correlations. 
Furthermore, initial conditions such as the color glass condensate 
(CGC) \cite{CGC} which invoke the coherent production of partons, might also lead to condensate formation \cite{CGCCondensate}.
In this article we present two- and three-pion correlations of 
same- and mixed-charge at low relative momentum to estimate the coherent fraction of charged-pion emission in Pb-Pb collisions 
at \snn= 2.76 TeV at the LHC with ALICE.

A number of past experimental efforts have been made to measure the degree of coherence in high-energy heavy-ion collisions using three-pion Bose-Einstein correlations: NA44, WA98, and STAR \cite{NA44Three,WA98Three,STARThree}.  
The methodology used here represents an improvement over the past efforts which we summarize in Sec.~$3$.

The remainder of this article is organized into 6 sections.  
In Sec.~$2$ we describe the data selection procedure.  
In Sec.~$3$ we introduce the methodology used in this analysis.
In Sec.~$4$ we describe the treatment of final-state interactions (FSIs).
In Sec.~$5$ we describe the treatment of momentum resolution corrections.
In Sec.~$6$ we explain the estimation of systematic uncertainties.
In Sec.~$7$ we present the results of this analysis. 
We conclude with a possible interpretation of the analysis results in Sec.~$8$.

\section{Experiment and data analysis}
Data were taken from the 2011 Pb-Pb run at \snn= 2.76 TeV at the CERN Large Hadron Collider (LHC) with ALICE \cite{ALICEdetector}.
The VZERO detectors \cite{VZERO}, located in the forward and backward regions of the detector, were used to form a minimum-bias trigger by requiring a simultaneous signal in both \cite{ALICEtrigger}.  The charged-particle multiplicity in the VZERO detectors is used to determine the collision centrality.
Approximately $34\times10^6$ minimum-bias collisions were used in this analysis.  Particle tracking was performed with two azimuthally complete detectors: the inner tracking system (ITS) and the time projection chamber (TPC) \cite{TPC}.  
The ITS consists of six layers of silicon detectors: silicon pixel (layers 1--2), silicon strip (layers 3--4), and silicon drift (layers 5--6) detectors.  The combined number of readout channels for all six layers is $1.257\times10^7$.
The ITS provides high spatial resolution to the distance of closest approach (DCA) of a particle to the primary vertex. However, it was not used for the momentum determination of particles in this analysis.  Cluster sharing within the ITS was found to cause a slight increase in track merging, to which this analysis is especially sensitive.
The TPC was used to determine the particle's momenta and charge via its radius of curvature in the 0.5-T longitudinal magnetic field.  The TPC is composed of 159 radially aligned pad rows for each of the 18 azimuthal sectors, totaling 557,568 readout channels. 

In addition to the tracking capabilities, the ITS and TPC provide particle identification capabilities through the specific ionization energy 
loss ($dE/dx$) in the silicon layers and TPC gas, respectively.  We select charged pions within 2 standard deviations ($\sigma$) of the expected pion $dE/dx$ value.  For momenta greater than 0.6 GeV/$c$, high pion purity is maintained with the time-of-flight (TOF) detector.  The TOF covers the full azimuthal 
range and the pseudo rapidity range $|\eta|<0.9$, except for the region $260^{\circ}<\varphi<320^{\circ}$ where no TOF modules were installed to reduce the material budget in front of the photon spectrometer.  With TOF we select tracks within 2 $\sigma$ of the expected pion TOF values.  
Tracks which are within 2 $\sigma$ of the expected kaon or proton
$dE/dx$ or TOF values are rejected.  Below 0.45 GeV/$c$ we further reject pion candidates if their $dE/dx$ is within 2 $\sigma$ of the expected electron $dE/dx$ value.  The pion-pair purity in this analysis is estimated to range from $90\%$ to $94\%$ for the highest and lowest pair momentum, respectively.

To ensure uniform tracking in the ITS, TPC, and TOF we require the $z$ coordinate of the primary vertex to be within a distance of 10 cm from the detector center.  We analyze 
tracks with transverse momenta in the interval $0.16<\pt<1.0$ GeV/$c$ and pseudorapidity $|\eta|<0.8$.  To ensure good momentum resolution, we require a minimum of 70 tracking points in the TPC. 

Track merging and splitting are known issues for same-charge tracks at very low relative momentum \cite{ALICEPbPbQSFirst}.  We minimize the contribution from merged and split pairs through three types of pair cuts.  First, we simply reject all pairs whose Lorentz invariant relative momentum, \qinv, is less than 5 MeV/$c$.  Second, we reject all pairs whose angular separation is less than 0.02 and 0.045 rad in the longitudinal and azimuthal direction, respectively.  The pair angular separation is evaluated at a radial distance of 1.0 and 1.6 m, where the most pronounced track-merging and -splitting effects were observed, respectively.  Third, we reject pairs that share more than $5\%$ of pad-row tracking points \cite{STARAuAuQS}.  These three cuts are applied to all terms of the correlation functions (same-event and mixed-event) introduced in the next section.  For three-pion correlations we apply these three cuts to each of the three pairs in the triplet.  The cuts are only applied to same-charge pairs.   Mixed-charge pairs are easily distinguished in the 
central barrel magnetic field as their trajectories are bent away from each other.

\section{Methodology}
Two-particle correlation functions are binned in narrow intervals of the mean pair transverse momentum, $\kt=|\bf{p_{\rm T,1}}+\bf{p_{\rm T,2}}|$/2, and Lorentz invariant relative momenta, \qinv~$=\sqrt{-(p_1-p_2)^{\mu}(p_1-p_2)_{\mu}}$.  They are defined as the ratio of the inclusive two-particle spectrum, $N_2(p_1,p_2)$ over the product of inclusive single-particle spectra, $N_1(p_1)N_1(p_2)$:
\begin{equation}
  C_2(p_1,p_2) = \frac{N_2(p_1,p_2)}{N_1(p_1)N_1(p_2)}.
  \label{eq:C2}
\end{equation}
The numerator of the correlation function is formed by all pairs of particles from the same event.  The denominator is formed by taking one particle from one event and the second particle from another event.  The same- and mixed-event two-particle distributions are normalized to each other in the interval $0.15<q<0.175$ GeV/$c$, sufficiently above the dominant region of low relative momentum correlations and sufficiently narrow to avoid the small influence of background correlations.  Only events within the same centrality class are mixed.  The centrality classes correspond to the 
top $0-5\%$ through $45-50\%$ of the particle multiplicity distribution estimated with the VZERO detector.  Each class has a width of $5\%$.  

The isolation of genuine two-pion correlations is complicated by several additional factors.
Namely, the resolvable threshold of low relative momentum pairs is limited by 
track merging and splitting in the ALICE detector.  The QS correlation of long-lived resonance decays is largely localized below 
this threshold and is therefore unobservable.  This leads to an apparent decrease of QS correlations and is described by 
the $\lambda$ or ``dilution'' parameter in this analysis. Given $\lambda$, two-particle correlations can be written as
\begin{eqnarray}
N_2(p_1,p_2) &=& {\mathcal N}[(1-\lambda)N_1(p_1)N_1(p_2) \nonumber \\
&+& \lambda K_2(\qinv)N_2^{QS}(p_1,p_2)], \label{eq:N2QS} \\
C_2(\qinv) &=& {\mathcal N}[(1-\lambda) + \lambda K_2(\qinv)C_2^{QS}(\qinv)],
\label{eq:C2QS}
\end{eqnarray}
where ${\mathcal N}$ is a residual normalization taking into account the small nonfemtoscopic contributions \cite{Bowler,Sinyukov}.  We allow a different ${\mathcal N}$ for same and mixed-charge correlations as the nonfemtoscopic contributions can be different.  $K_2(\qinv)$ is the FSI correlation.  $N_2^{QS}$ and $C_2^{QS}(\qinv)$ are the genuine two-pion QS distribution and correlation, respectively.  Here, unlike in most experimental publications on this subject, the $\lambda$ parameter does not include effects of partial 
coherence.  Its deviation below unity can also be attributable to secondary contamination, pion misidentification, and finite \qinv~binning.
Same-charge pion QS correlations excluding coherence can be parametrized by
\begin{eqnarray}
C_2^{QS,++}(\qinv) &=& 1 + E_{\rm w}(R_{ch}\qinv)^2e^{-R_{ch}^2\qinv^2}, \label{eq:C2ssparameters} \\
E_{\rm w}(R_{ch}\qinv) &=& 1 + \sum_{n=3}^{\infty} \frac{\kappa_n}{n! (\sqrt{2})^n} H_n(R_{ch}\qinv),
\end{eqnarray}
where $R_{ch}$ are the characteristic radii of the chaotic 
component.  $E_{\rm w}(R_{ch}\qinv)$ is the Edgeworth expansion characterizing deviations from Gaussian behavior \cite{TamasEW}.  $H_n$ are the Hermite polynomials and $\kappa_n$ are the Edgeworth coefficients.  The first two relevant Edgeworth coefficients
($\kappa_3, \kappa_4$) are found to be sufficient to describe the non-Gaussian features in this analysis.  At the two-pion level we do not include 
an explicit parametrization of a possible coherent component owing to the large uncertainty of non-Gaussian Bose-Einstein correlations.
In this analysis we assume $\lambda$ of mixed-charge pions is identical to that of same-charge pions: $\lambda^{+-}=\lambda^{\pm\pm}$.  This is a valid assumption at high energies where the secondary contamination from particles and antiparticles are expected to be equal \cite{ALS}.

Three-particle correlation functions are binned in terms of the three invariant relative momenta in the triplet: $q_{12}$, $q_{31}$, and $q_{23}$.
The three-particle correlation function is similarly the ratio of the inclusive three-particle spectrum to the product of the inclusive single-particle spectra binned in the pair relative momenta:
\begin{eqnarray}
  C_3(p_1,p_2,p_3) &=& \frac{N_3(p_1,p_2,p_3)}{N_1(p_1)N_1(p_2)N_1(p_3)}, \label{eq:C3} \\
  Q_3 &=& \sqrt{q_{12}^2+q_{31}^2+q_{23}^2}. \label{eq:Q3}
\end{eqnarray}
The numerator of $C_3$ is formed by all triplets of particles from the same event.  The denominator is formed by taking each of the three particles from different events. 
We project three-particle correlations against the Lorentz invariant $Q_3$.

For three-particle correlations, $\lambda \neq 1$ similarly causes ``feed-up'' from pure combinatorial distributions and two-particle correlations as described in Eq.~(\ref{eq:N3QS}) below.  The derivation of Eq.~(\ref{eq:N3QS}) is shown in the Appendix.
In Eq.~(\ref{eq:N3QS}), $N_2(p_{\rm i},p_{\rm j})N_1(p_{\rm k})$ terms represent the case where particles $i$ and $j$ are taken from the same event while particle $k$ is taken from a different 
event and $K_3$ is the three-pion FSI correlation.  Isolation of the three-pion QS correlation is done by solving Eq.~(\ref{eq:N3QS}) for $N_3^{QS}$.  Using $N_2^{QS}$ and 
$N_3^{QS}$ one can construct a cumulant correlation function, ${\rm {\bf c}_3}$, in Eq.~(\ref{eq:c3}):
\begin{eqnarray}
N_3(p_1,p_2,p_3) &=& f_1N_1(p_1)N_1(p_2)N_1(p_3) \nonumber \\
&+& f_2\big[N_2(p_1,p_2)N_1(p_3) + N_2(p_3,p_1)N_1(p_2) + N_2(p_2,p_3)N_1(p_1) \big] \nonumber \\
&+& f_3K_3(q_{12}, q_{31}, q_{23})N_3^{QS}(p_1,p_2,p_3) \label{eq:N3QS} \\ 
{\rm {\bf c}_3}(p_1,p_2,p_3) &=& 1 + \big[2N_1(p_1)N_1(p_2)N_1(p_3) \nonumber \\
&-& N_2^{QS}(p_1,p_2)N_1(p_3)-N_2^{QS}(p_3,p_1)N_1(p_2) - N_2^{QS}(p_2,p_3)N_1(p_1) \nonumber \\
&+& N_3^{QS}(p_1,p_2,p_3)\big]/N_1(p_1)N_1(p_2)N_1(p_3).
\label{eq:c3}
\end{eqnarray}
\noindent In Eq.~(\ref{eq:N3QS}), $f_1$, $f_2$, and $f_3$ are derived in the Appendix and are given by $(1-\lambda^{1/2})^3+3\lambda^{1/2}(1-\lambda^{1/2})^2 - 3(1-\lambda^{1/2})(1-\lambda)$, $(1-\lambda^{1/2})$, $\lambda^{3/2}$, respectively.

The quantity in square brackets in Eq.~(\ref{eq:c3}) represents a three-pion cumulant which has all two-pion correlations removed.  Therefore, 
the three-pion cumulant represents the isolation of genuine three-pion QS correlations.
All same and mixed-event three-particle distributions are normalized to each other in the range where all three pairs satisfy $0.15<q_{ij}<0.175$ GeV/$c$, sufficiently above the dominant region of low relative momentum correlations and sufficiently narrow to avoid the small influence of background correlations.

The novel effects measured with three-particle correlations are isolated with the $r_3$ function \cite{UliChaoticity,UliProjection}:  
\begin{equation}
r_3(p_1,p_2,p_3) = \frac{{\rm {\bf c}_3}(p_1,p_2,p_3) - 1}{\sqrt{(C_2^{QS}(p_1,p_2)-1)(C_2^{QS}(p_3,p_1)-1)(C_2^{QS}(p_2,p_3)-1)}}.
\label{eq:r3}
\end{equation}
The $r_3$ function isolates the phase of three-pion correlations: $r_3 = I \cos(\Phi)\approx I(1 - \Phi^2/2)$ \cite{UliChaoticity}.  The intercept of $r_3$, $I$, is expected to be 2 in the case of fully chaotic particle-emitting sources and less than 2 in the case of partially coherent 
sources.  The leading-order contribution to the phase was shown to be quadratic in relative momenta, $\Phi \approx a_{\mu\nu}q^{\mu}_{12}q^{\nu}_{23}$, which leads to quartic behavior in $r_3$ \cite{UliChaoticity}.  The antisymmetric tensor $a_{\mu\nu}$ characterizes space and momentum source asymmetries related to how the spatial position of maximum pion emission changes with momentum.  There are six nonvanishing independent components in $a_{\mu\nu}$.  However, owing to limited statistical precision we project $r_3$ from three-dimensional invariant relative momenta to one-dimensional $Q_3$.  A fit quartic and quadratic in $Q_3$ is performed,
\begin{eqnarray}
r_3(Q_3) = I(1-aQ_3^4), \label{eq:QuarticFit} \\
r_3(Q_3) = I(1-aQ_3^2), \label{eq:QuadraticFit}
\end{eqnarray}
where $I$ is the intercept of $r_3$ ($I=r_3(0)$), and $a$ is the quartic or quadratic coefficient.  The quadratic fit is motivated by previous 
fit attempts by the STAR collaboration \cite{STARThree}.  
The coherent fraction ($G$) can be extracted from the intercept as \cite{UliChaoticity}
\begin{equation}
I = 2\sqrt{1-G} \frac{1+2G}{(1+G)^{3/2}}.
\label{eq:Gfromr3}
\end{equation}
Equation (\ref{eq:Gfromr3}) neglects the effect of the charge constraint on charged coherent states \cite{GenCohStates,ThreeGenCoh,ALS}.  In the quantum optics approach to coherent 
states \cite{Glauber}, charged pions can only be in coherent states when positive and negative pions pair together to form a charge neutral state.  
However, because the charge constraint affects both numerator and denominator of $r_3$ in the same direction, its effect on $r_3$ for $G<30\%$ is expected to increase its intercept by less than $17\%$ \cite{ThreeGenCoh}.

The denominator of $r_3$ is measured using the three-particle combinatorial distribution and two-particle correlation strengths.  The two-particle correlation 
strengths are tabulated from a previous run over the data.  They are tabulated in sufficiently narrow intervals or bins of centrality, \kt, and three-dimensional relative momentum to allow reliable interpolation between bins. 
We bin the two-particle correlations in nine centrality bins ($5\%$ wide) and four \kt~bins in the longitudinally comoving system (LCMS).  Forty $q_{out}$, $q_{side}$, and $q_{long}$ bins (5 MeV/$c$ wide) are chosen.  
$q_{out}$ is the projection of the relative momentum along the pair momentum direction.  $q_{long}$ is the projection along the beamline.  $q_{side}$ is then perpendicular to the other two (azimuthal projection).  
The four \kt~bins are chosen such that they divide the pair distribution into four equally populated intervals.

\subsection{Methodology Improvement}
The methodology used here to measure three-pion QS correlations represents an improvement over the past efforts \cite{NA44Three,WA98Three,STARThree}, which we highlight here.
\begin{enumerate}
\item In addition to QS correlations, charged pions also experience a Coulomb repulsion which reduces the apparent strength of QS correlations.   
Corrections for the three-body Coulomb interactions are damped in this analysis according to the observed $\lambda$ parameter.  Previously, the 
Coulomb corrections were undamped and thus overestimated.  
\item The Coulomb corrections are estimated by integrating over an assumed freeze-out distribution of pions.  We take into account the effect of resonance decays on the freeze-out distribution.  Previously, a Gaussian distribution was assumed.
\item For the case when $\lambda<1$, the measured three-pion correlations contain a feed-up from lower-order correlations, which is now removed.
\item We apply momentum resolution corrections, which was not universally done in the past efforts.
\item We apply corrections for muon contamination which was not done in the past efforts.
\item The isolation of the cumulants is done at the pair/triplet distribution level instead of at the correlation 
function level.
\item Mixed-charge two- and three-pion correlations are used to help determine the $\lambda$ parameter and to monitor the performance of FSI corrections.
\end{enumerate}

\section{Final-State-Interactions}
The treatment of FSIs is crucial for this analysis.  
In addition to QS correlations, identical charged pions also experience FSIs which reduce the apparent strength of QS correlations.  
The FSIs of charged pions are dominated by the Coulomb interaction.  
The strong interactions, while small for same-charge pions, are important for mixed-charge pions.  Coulomb and strong FSI corrections are included in this analysis for both two- and three-particle same- and mixed-charge correlations.  The wave functions for 
two-pion Coulomb and strong FSIs are known to high precision \cite{LednickyFSI}.  Two-pion FSIs are calculated by averaging the modulus square of the  
two-pion FSI wave functions over an assumed freeze-out particle-emitting source distribution.  This is then divided by the corresponding average of plane-wave functions to 
isolate the pure FSIs.  For same-charge pions, the wave functions are symmetrized.
Typically the source distribution is taken to be a spherical Gaussian with 
a radius matching what is found in the data.  Here, we use a more sophisticated approach.  All FSIs are calculated directly within \textsc{therminator} 2 events \cite{Therminator, ThermBozek}.  The pair relative separation at freeze-out in the pair-rest frame, $r^*$, as well 
as the space-momentum correlations included in the model are used.  
\textsc{therminator} includes all of the known resonance decays.  Pions from resonance decays add non-Gaussian features to the freeze-out distribution.  Furthermore, they increase the mean value of $r^*$, which in turn reduces the strength of FSI correlations.  The same centrality class and \kt~range from the data are used to calculate the FSIs.  The freeze-out hyper-surfaces in \textsc{therminator} were calculated within 3D viscous hydrodynamics with an 
initial and final temperature of 512 and 140 MeV, respectively.  The starting time for hydrodynamics was 0.6 fm/$c$.

Three-body FSI wave functions are not known for all regions of phase-space.  However, all asymptotic wave functions are known \cite{Alt3body}.  In particular, the wave-function 
corresponding to the phase-space region where all three inter particle spacings are large, $\Omega_0$, is given by the product of the three two-body 
wave functions.  It has been shown 
that the $\Omega_0$ wave function is a justified approximation also in the case where the triplet kinetic energy in the triplet rest frame is sufficiently 
large \cite{CsorgoOmega0}.  It is estimated that triplet energies exceeding about 7 MeV for 6-fm sources justify the use of the $\Omega_0$ wave function.  The minimum triplet energy considered in this analysis is $\sqrt{3}\times 5\approx 8.7$ MeV when all three pair \qinv's~are at their minimum allowed value of 5 MeV/$c$.  

For the case of same-charge pion FSIs with the $\Omega_0$ wave function, the modulus square of the fully symmetrized FSI wave-function is averaged in \textsc{therminator} events.  This is then 
divided by the corresponding average of fully symmetrized plane waves.  The full symmetrization assumes fully chaotic emission.  For the case of mixed-charge FSIs, only the same-charge pairs are symmetrized.  All $K$ factors in this analysis are averaged over the \textsc{therminator} 
freeze-out distribution for pairs satisfying $r^*<80$ fm.  For the $K_3$ calculation, all three pairs must satisfy this requirement.

All three-pion correlations in this analysis are binned in 3D corresponding to the three pair invariant 
relative momenta: $q_{12}$, $q_{23}$, $q_{31}$.  The three-pion FSI correlations are likewise calculated in 3D for the integrated \kt~range.

\begin{figure}
  \center
  \includegraphics[width=0.49\textwidth]{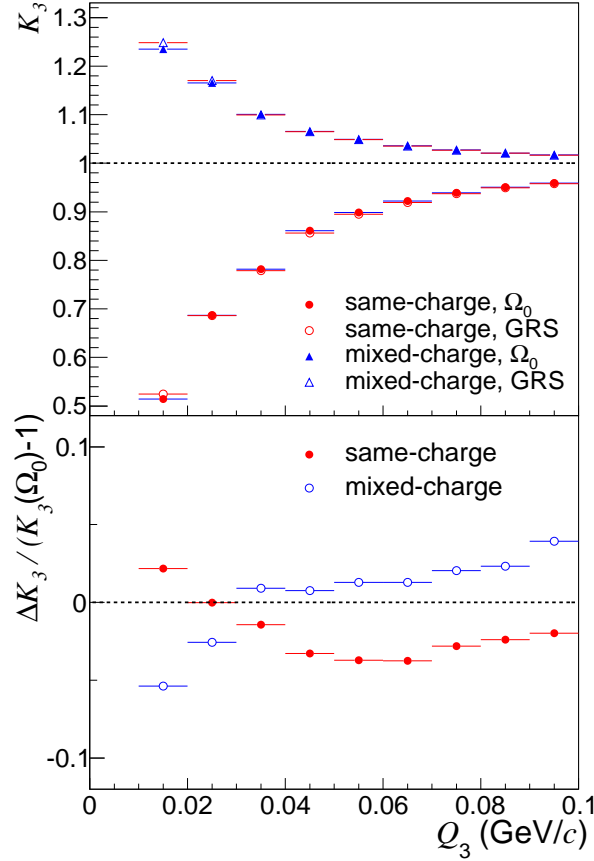}
  \caption{Comparison of same and mixed-charge three-pion FSI correlations.  $\Omega_0$ wave-function and generalized Riverside (GRS) method are shown. 
The calculation was performed in \textsc{therminator} ($0-5\%$).  The bottom panel
    shows the difference between the two methods, $\Delta K_{3}=K_{3}(\Omega_{0})-K_{3}(GRS)$, divided by $K_{3}(\Omega_{0})-1$.}
  \label{fig:K3Omega0GRS}
\end{figure}

Another more commonly used approach to treat three-body FSIs is the Riverside approach \cite{Riverside} for which the three-body FSI correlation, 
$K_3$, is given by the triple product of Gamov factors ($K_3 = G^{12}G^{23}G^{31}$).  In the generalized version of this approach, ``generalized Riverside" (GRS), each two-body factor is 
averaged over the assumed source distribution ($K_3 = K_2^{12}K_2^{23}K_2^{31}$) \cite{WA98Three, STARThree}.  In Fig.~\ref{fig:K3Omega0GRS} we compare our calculations of three-body FSI correlations using the $\Omega_0$ wave function and GRS approach within \textsc{therminator} events.  
We observe similar FSI correlations with both methods. 

\section{Momentum Resolution}
Finite momentum resolution in the ALICE detector generally causes a smearing of the correlation function.  We estimate its effect on the correlation functions by assigning a weight to each pair or triplet 
in \textsc{hijing} \cite{HIJING} based on the measured correlation strength in real data.  The same weight is applied to two versions of each $N_n$ ($n=1,2,3$) histogram.  The first is filled with the nonsmeared ideal \qinv~from \textsc{hijing}.  The second 
is filled with the smeared \qinv~after the tracks have been propagated through the simulation of the ALICE detector response.  The ratio of the first to the second histogram forms the correction factor for the $N_n$ distributions.  

The momentum resolution corrections are found to be largest at low \qinv~($Q_3$), where they increase the raw correlation function by less than $5\%$ ($8\%$) for two-pion (three-pion) correlations.  We also observe that the correction factors do not change significantly with \kt.
After the momentum resolution corrections are 
applied, we verified that the observed correlation strength and shape matches the assumed values used as a weight in \textsc{hijing}.

\section{Muon Contamination}
The pion-pair purity is estimated to be about $93\%$ in \textsc{hijing} with the simulated ALICE detector response.  
The leading order misidentified pair is the muon pion combination.  
The rest of the misidentified combinations taken together contribute less than $1\%$ to the total pairs.
We estimate that about $93\%$ of the muons contaminating our sample originate from primary-pion decays.
The primary parent pion is expected to interact with the other primary pions via QS+FSI.  
We therefore expect that the muon pion pairs contaminating our sample will contain a residual pion pion correlation.  
For the three-pion case the muon pion pion combination dominants the misidentified triplets.
We form a correction factor for all two-pion (three-pion) terms by assigning a QS+FSI weight to the parent pions in the pair (triplet) which subsequently decayed into muons.  
A smeared correlation is obtained when the assigned correlation is binned in relative momentum using the muon momentum.
The ratio of the assigned correlation to the smeared correlation forms our correction factor.  
The correction is applied to same and mixed-charge correlations and is found to increase $\lambda$ by about $5\%$ while having a negligible effect on the extracted radii.  The correction increases the two-pion correlation by about $1.5\%$ at low $q$ and rapidly decreases for larger $q$.  The correction increases the three-pion correlation by about $3\%$ at low $Q_3$ and by about $1\%$ for high $Q_3$.

\section{Systematic Uncertainties}
The dominant systematic uncertainty in this analysis pertains to the unknown spatio temporal pion distribution at freeze-out on which the fitting of the correlation functions and FSI calculations depends.  
Typically, a Gaussian profile is assumed in most femtoscopic analyses.  However, the known resonances taken all together will generally give rise to non-Gaussian features in the freeze-out distribution.  

The systematic uncertainty of the freeze-out distribution is two fold in this analysis. 
First, it creates an uncertainty in the wave-function integration for the FSI calculation. 
However, the $q$ dependence of FSI correlations is largely invariant to reasonable variations of the assumed freeze-out distribution and radius.  
A possible mismatch of the freeze-out distribution and radius in \textsc{therminator} as compared to the data is largely absorbed by the $\lambda$ parameter of the global fits to same- and mixed-charge two-pion correlations presented in the 
Results section.  We assign a $2\%$ uncertainty on the two-pion FSI correlations based on the maximum observed difference between FSIs calculated in \textsc{therminator} and Gaussian particle-emitting source profiles after rescaling by an effective $\lambda$ parameter. 
We also assign a $2\%$ uncertainty on the $r^*$-dependent part of the FSI wave functions \cite{LednickyFSI}.
Second, the freeze-out distribution uncertainty creates an uncertainty in the fitting of the same-charge correlation functions.  A convenient account of sufficiently small deviations from Gaussian behavior in the QS correlation functions can be obtained through an 
Edgeworth expansion \cite{TamasEW}.  Deviations from Gaussian behavior are also expected from a finite coherent component \cite{ALS}.

Non-Gaussian features in the QS correlation functions can also occur in more trivial ways.
Spherical Gaussian freeze-out distributions create Gaussian QS correlation functions 
as a function of \qinv.  Non-Gaussian features in 1D correlation functions can arise simply from nonequal 3D radii in the LCMS frame.  However, we note that $R_{out} \approx R_{side}$ 
and $R_{long}$ is only $20\%$ larger than $R_{out}$ and $R_{side}$ \cite{ALICEPbPbQSFirst}.  Also, \kt~and centrality bins whose widths are not sufficiently narrow will create a mix of different radii and therefore will  
not be described by a single Gaussian function.  However, our chosen centrality bin width ($5\%$) and \kt~bin width (100 MeV/$c$ for two-particle correlations) are sufficiently narrow to mostly avoid this feature given the known 
\kt~dependencies of the radii \cite{ALICEPbPbQSFirst}.  More non-Gaussian features are expected for our three-particle correlations as the \kt~bin is much wider (1 GeV/$c$).  

The momentum resolution of low-momentum particles ($\pt<1$ GeV/$c$) is dominated by multiple scatterings within the ALICE detector.  The ALICE 
material budget uncertainty is conservatively estimated to be $\pm10\%$.  Our studies suggest a near one-to-one correspondence of the material 
budget uncertainty with the momentum resolution uncertainty.  We apply a $10\%$ uncertainty on all the momentum resolution corrections.  
For $r_3$ the momentum resolution correction uncertainty is found to be $1\%$.  It is not the dominant uncertainty since both numerator 
and denominator are affected in the same direction.

We study the uncertainties associated with tracking in the ALICE detector in several ways.  We study the effect of different magnetic-field 
orientations in the TPC.  The pion particle identification (PID) cuts are tightened by $10\%$.  The angular separation cuts for same-charge pairs are increased by $50\%$.  Positive pions are compared to negative pions.  All the uncertainties in this category except for PID were found to be negligible.  
A $0.3\%$ and $1\%$ systematic uncertainty owing to PID were assigned for three-pion correlation functions and $r_3$, respectively.

Concerning $r_3$, additional systematics are included.
Imperfect isolation of the three-pion QS cumulant (FSI corrected) is the dominant uncertainty for $r_3$ which mostly affects the larger values 
of $Q_3$ where the cumulant is smallest.
The chosen $\lambda$ parameter ($\lambda=0.7$) used in extracting the QS correlations in both the numerator and the denominator, while largely canceling in the ratio, is varied by $0.1$.  Mixed-charge three-pion cumulant correlations (${\rm {\bf c}_3}^{\pm\pm\mp}$) reveal a slight residual correlation of about
$1.005$ for all centralities.  The residual cumulant correlation in the mixed-charge channel is used as a systematic uncertainty in the same-charge channel.  Also, small variations of the powers $m$ and $n$ in Eq.~(\ref{eq:N3QS}) 
which brought ${\rm {\bf c}_3}^{\pm\pm\mp}$ closer to unity resulted in similar systematic variations for $r_3$.
This procedure is valid if the true FSI corrected mixed-charge cumulant correlation is expected to be near unity.

The GRS approach to Coulomb 
corrections is found to give a better description of the mixed-charge correlations than the $\Omega_0$ wave function.  For this reason we choose the GRS approach as 
our principal method and use the $\Omega_0$ wave function as a systematic variation for all three-pion correlations.  
Finally, nonfemtoscopic background correlations associated with minijets \cite{ALICEpp}, while negligible for the highest multiplicity collisions, create a small uncertainty in the extraction of 
two-pion QS correlation strengths.  A linear fit to the background is made in the interval $0.2<q<0.4$ GeV/$c$ and extrapolated into 
the femtoscopic region, $q<0.15$ GeV/$c$.  The correction only has a non-negligible effect on $r_3$ for large $Q_3$ and above $40\%$ centrality.

\section{Results}
\subsection{Two Pions}
\begin{figure}
  \subfigure[$0-5\%$ centrality.]{
  \includegraphics[width=0.49\textwidth]{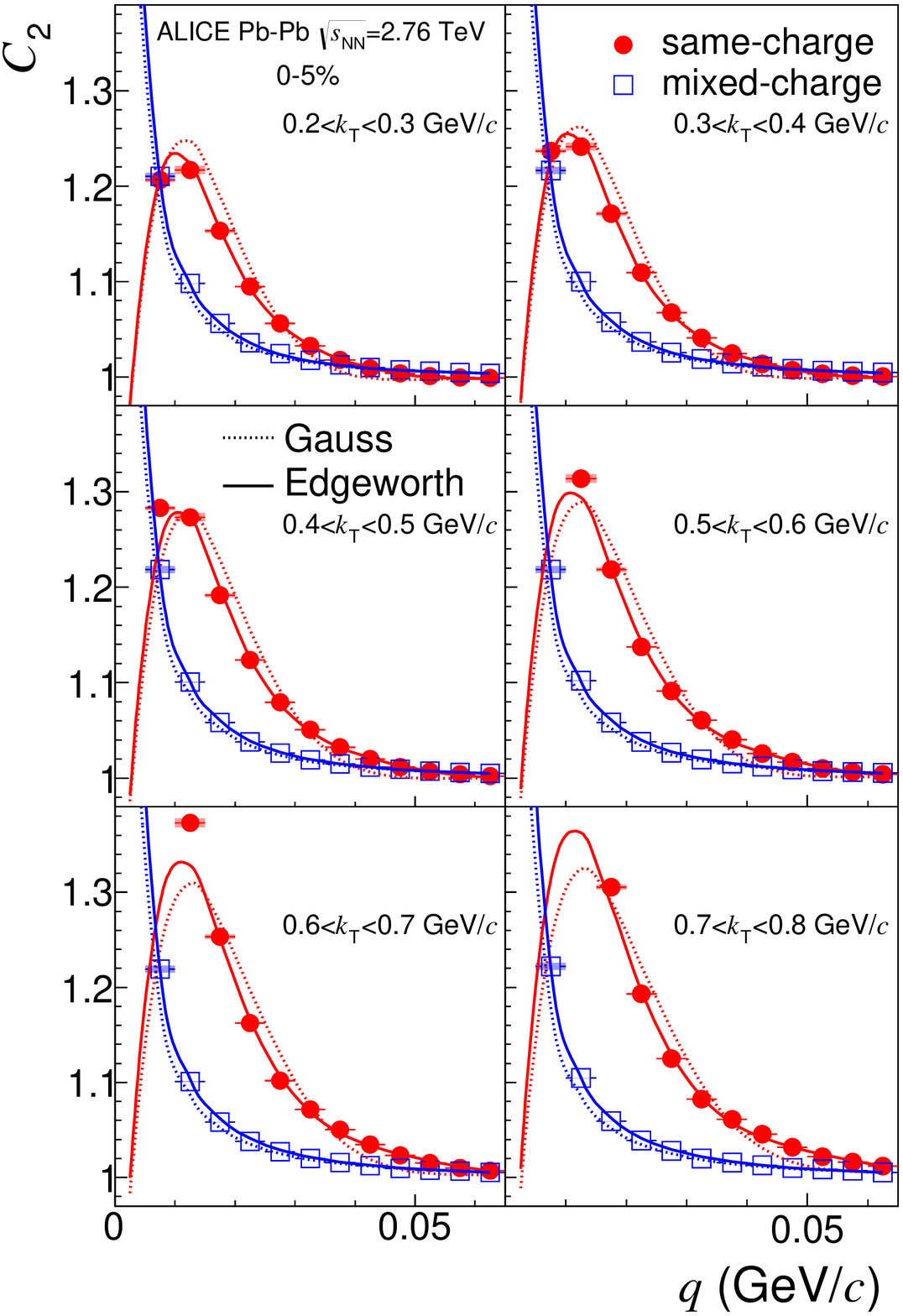}
\label{fig:C2globalfitM0}
  }
  \subfigure[$45-50\%$ centrality.]{
    \includegraphics[width=0.49\textwidth]{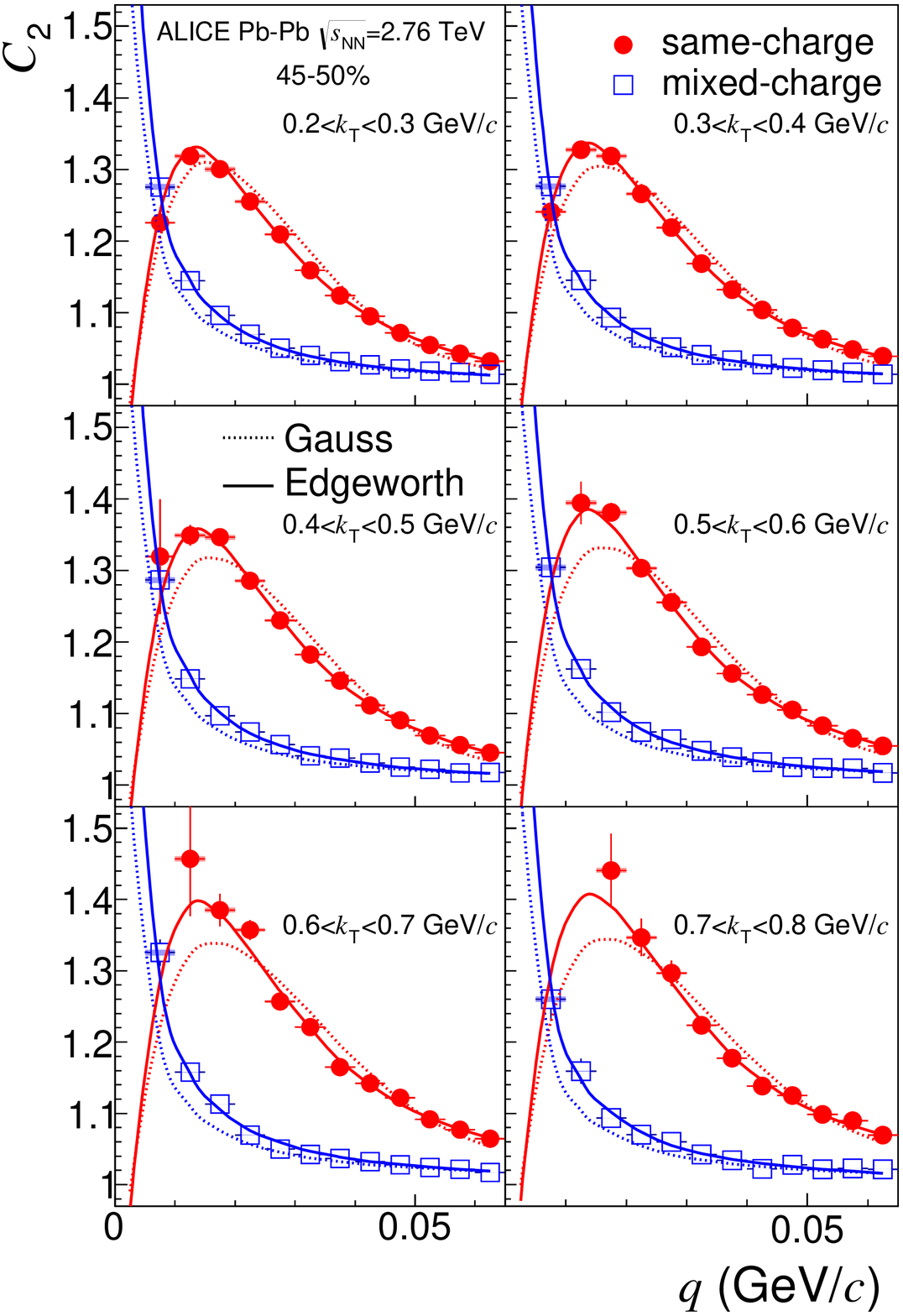}
    \label{fig:C2globalfitM9}
  }
  \caption{$C_2$ for same-charge (solid red circles) and mixed-charge pions (open blue squares) for $0-5\%$ centrality (a) and $45-5-\%$ centrality (b).  
The global fits with dotted lines correspond to Gaussian same-charge fits ($E_{\rm w}=1$).  The global fits with solid lines correspond to non-Gaussian fits with Edgeworth coefficients ($E_{\rm w}\neq1$).  Shaded boxes represent the momentum resolution correction uncertainty.  FSI uncertainties are smaller than the symbol sizes.}
\end{figure}

We first present the two-pion correlation functions.  Figures \ref{fig:C2globalfitM0} and \ref{fig:C2globalfitM9} show the same- and mixed-charge correlation functions versus \qinv~in 6 \kt~bins for $0-5\%$ and $45-50\%$ centrality, respectively.  Global fits for same and mixed-charge 
correlations are performed for each \kt~bin separately.  Two types of global fits are shown.  The dotted lines correspond to Gaussian fits ($E_{\rm w}=1$), while the solid lines correspond to non-Gaussian fits with Edgeworth coefficients ($E_{\rm w}\neq1$).  
Our strict pair cuts cause a lack of data for same-charge correlations at low \qinv~at high \kt~where a larger fraction of the pairs moves collinearly and 
thus is more susceptible to track merging and splitting.

Concerning the purely Gaussian fits in Figs.~\ref{fig:C2globalfitM0} and \ref{fig:C2globalfitM9}, the average $\chi^2$ per degree of freedom ($NDF$) is 39.  It is clear 
that a spherical Gaussian fully chaotic source can be ruled out.  The global fits underestimate mixed-charge correlations for each \kt~and centrality bin.  
The fits indicate the possibility of significant non-Gaussian features in the same-charge correlation functions and/or the possibility of two 
separate suppression parameters.  An individual fit to mixed-charge correlations suggests $\lambda$ is about 0.7.  An individual fit to 
same-charge correlations with a Gaussian function suggests a value of about 0.4. 

Concerning the Edgeworth fits in Figs.~\ref{fig:C2globalfitM0} and \ref{fig:C2globalfitM9}, the average $\chi^2/NDF$ is 1.5.  Same- and mixed-charge 
correlations are simultaneously well described with an Edgeworth fit.  A common $\lambda$ parameter is now able to describe both same- and mixed-charge correlations.  
This may demonstrate the significance of non-Gaussian same-charge correlations and/or the presence of a coherent component. 

Fits including coherence with and without the charge constraint were also attempted.  
The charge constraint on coherent states in the quantum optics \cite{Glauber} approach leads to a slight modification of both same-charge 
and mixed-charge correlations \cite{ALS}.  It leads to a slight decrease of the suppression of same-charge correlations 
($\frac{1}{5}G^2$) and also an enhancement of mixed-charge correlations ($\frac{1}{5}G^2$) \cite{ALS}.
Coherence may also 
explain the observation of separate suppression parameters as it only suppresses same-charge 
correlations.  However, given the uncertainty of non-Gaussian same-charge correlations, 
we find that two-pion correlations alone are inconclusive in determining the 
presence of coherence.

\begin{figure}
\center
  \includegraphics[width=0.49\textwidth]{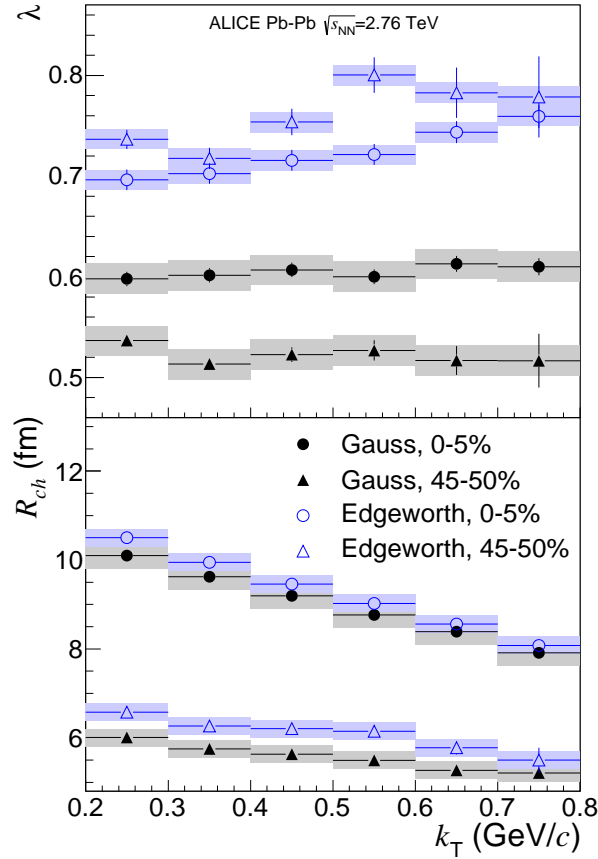}
  \caption{Fit parameters versus \kt~for Gaussian and Edgeworth global fits in Figs.~\ref{fig:C2globalfitM0} and \ref{fig:C2globalfitM9}. 
(Top) $\lambda$ values.  (Bottom) $R_{ch}$ values.  Shaded bands represent systematic uncertainties.}
  \label{fig:C2fitParams}
\end{figure}

The $\lambda$ and radii fit parameters for both global fit types are shown in Fig.~\ref{fig:C2fitParams}. 
The Edgeworth coefficients from ALICE data are shown in Table \ref{tab:ALICEkappa}.
\begin{table}
  \center
  \begin{tabular}{| c | c | c | c | c | c | c | c |}
    \hline
    $\kappa_3$ & $\kt_1$ & $\kt_2$ & $\kt_3$ & $\kt_4$ & $\kt_5$ & $\kt_6$ \\ \hline
    $0-5\%$ & 0.14 & 0.13 & 0.12 & 0.12 & 0.1 & 0.094  \\ \hline
    $45-50\%$ & 0.23 & 0.22 & 0.23 & 0.25 & 0.25 & 0.24  \\ \hline
    $\kappa_4$ & & & & & & \\ \hline
    $0-5\%$ & 0.29 & 0.33 & 0.37 & 0.38 & 0.43 & 0.46  \\ \hline
    $45-50\%$ & 0.19 & 0.22 & 0.22 & 0.24 & 0.25 & 0.31  \\ \hline
  \end{tabular}
  \caption{$\kappa_3$ and $\kappa_4$ Edgeworth coefficients from ALICE data corresponding to 
global fits in Figs.~\ref{fig:C2globalfitM0} and \ref{fig:C2globalfitM9}.  $\kt_1$ and $\kt_6$ represent our lowest and highest $\kt$ intervals, respectively.}
  \label{tab:ALICEkappa}
\end{table}
The corresponding Edgeworth coefficients from \textsc{therminator} are shown in Table \ref{tab:Thermkappa}.
\begin{table}
  \center
  \begin{tabular}{| c | c | c | c | c | c | c | c |}
    \hline
    $\kappa_3$ & $\kt_1$ & $\kt_2$ & $\kt_3$ & $\kt_4$ & $\kt_5$ & $\kt_6$ \\ \hline
    $0-5\%$ & 0.18 & 0.22 & 0.27 & 0.31 & 0.35 & 0.4  \\ \hline
    $45-50\%$ & 0.25 & 0.27 & 0.3 & 0.34 & 0.36 & 0.42  \\ \hline
    $\kappa_4$ & & & & & & \\ \hline
    $0-5\%$ & 0.076 & 0.12 & 0.17 & 0.18 & 0.22 & 0.23  \\ \hline
    $45-50\%$ & 0.034 & 0.061 & 0.081 & 0.085 & 0.11 & 0.084  \\ \hline
  \end{tabular}
  \caption{$\kappa_3$ and $\kappa_4$ Edgeworth coefficients from \textsc{therminator}.  $\kt_1$ and $\kt_6$ represent our lowest and highest $\kt$ intervals, respectively.}
  \label{tab:Thermkappa}
\end{table}
The Edgeworth coefficients presented in Tables \ref{tab:ALICEkappa} and \ref{tab:Thermkappa} quantify the non-Gaussian structure of the same-charge 
correlation functions.  They may also be influenced by a coherent component.  The comparison of Table \ref{tab:ALICEkappa} to Table \ref{tab:Thermkappa} demonstrates a discrepancy in the shape of QS correlations between \textsc{therminator} and ALICE data.

The values for the overall normalization, ${\mathcal N}$, are typically within 0.005 from unity.
We observe that $\lambda$ is about 0.7 and is largely \kt~independent for the Edgeworth fits.  
The pion-pair purity and the primary-pair purity in this 
analysis are estimated to be about $93\%$ and $84\%$, respectively.  
The correction for muon contamination accounts for pion misidentification.  We therefore expect $\lambda<0.84$.
The Gaussian radii are larger than what is typically reported \cite{ALICEPbPbQSFirst} owing to the global fit procedure which 
incorporates mixed-charge correlations to better constrain the $\lambda$ parameter.
The Edgeworth radii for the chaotic 
component are observed to be larger than the purely Gaussian radii by about $10\%$.  
We note that it has also been shown that the presence of a finite coherent component can influence the 
width ($\propto 1/R_{ch}$) of same-charge correlations \cite{WeinerCoherence,Gyulassy,ALS}.  In particular, for the case when the radius of a 
coherent component is smaller than the chaotic component same-charge correlations appear broader than expected by the chaotic component 
alone.  This can incorrectly give the impression of a smaller chaotic source.  This may also arise from a momentum dependence of a coherent component (not considered in our fits).  For all cases, we observe $R_{ch}$ to decrease with increasing \kt.  

A comparison of the \kt~evolution of same- and mixed-charge correlations in Figs.~\ref{fig:C2globalfitM0} and \ref{fig:C2globalfitM9} reveals that same-charge correlations change rapidly with increasing \kt~while 
mixed-charge correlations change very little.  The widening of same-charge correlations with increasing \kt~is potentially caused by radial flow 
\cite{PrattKt,SinyukovKt}.  In an expanding source, pairs with large \kt~are preferentially formed from particles within the same space-time interval.  Thus, 
larger values of \kt~measure smaller lengths of homogeneity.  In QS correlations, this will demonstrate itself as a widening of the correlation 
function with increasing \kt.  

\begin{figure}
  \center
  \includegraphics[width=0.49\textwidth]{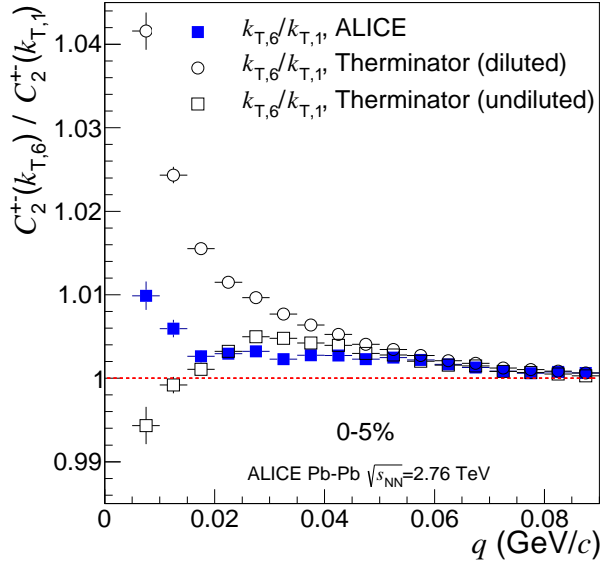}
  \caption{Ratio $C_2^{+-}(\kt_6) / C_2^{+-}(\kt_1)$, comparing mixed-charge correlations between the highest (sixth) and lowest (first) \kt~bins.  Open circles represent the \textsc{therminator} comparison using 
all pion pairs (diluted).  Open squares represent the \textsc{therminator} calculation only using pion pairs with $r^*<80$ fm (undiluted).  Error bars include statistical and systematic uncertainties.}
  \label{fig:Dpm}
\end{figure}

Similarly, mixed-charge pairs of larger \kt~may also measure smaller lengths of homogeneity owing to radial flow.  Mixed-charge 
correlation strengths may therefore increase with increasing \kt~because FSI correlations are larger for smaller sources.
In Fig.~\ref{fig:Dpm} we present mixed-charge correlations in the form of a ratio, $C_2^{+-}(\kt_6) / C_2^{+-}(\kt_1)$,
where $\kt_6$ and $\kt_1$ represent our highest (sixth) and lowest (first) \kt bins, respectively.
Comparing the ALICE data to the diluted \textsc{therminator} calculation in Fig.~\ref{fig:Dpm}, it is clear that the observed mixed-charge correlations evolve less rapidly in real data as compared to the \textsc{therminator} expectation.  
This may be caused by a discrepancy of $\lambda$ or the freeze-out size in \textsc{therminator} as compared to the data.  To distinguish between them, we also compare 
the ALICE data to the undiluted \textsc{therminator} calculation in Fig.~\ref{fig:Dpm} where only ``interacting" pairs with $r^*<80$ fm are used.  Such a procedure can help remove the effect of the $\lambda$ parameter from the comparison.  The \kt~evolution of mixed-charge correlations is better described with the 
undiluted \textsc{therminator} expectation which indicates a discrepancy of the \kt~evolution of the $\lambda$ parameter in \textsc{therminator} as 
compared to the data.

\subsection{Three Pions}

\begin{figure}
  \subfigure[$0.16<K_{\rm T,3}<0.3$ GeV/$c$.]{
    \includegraphics[width=0.49\textwidth]{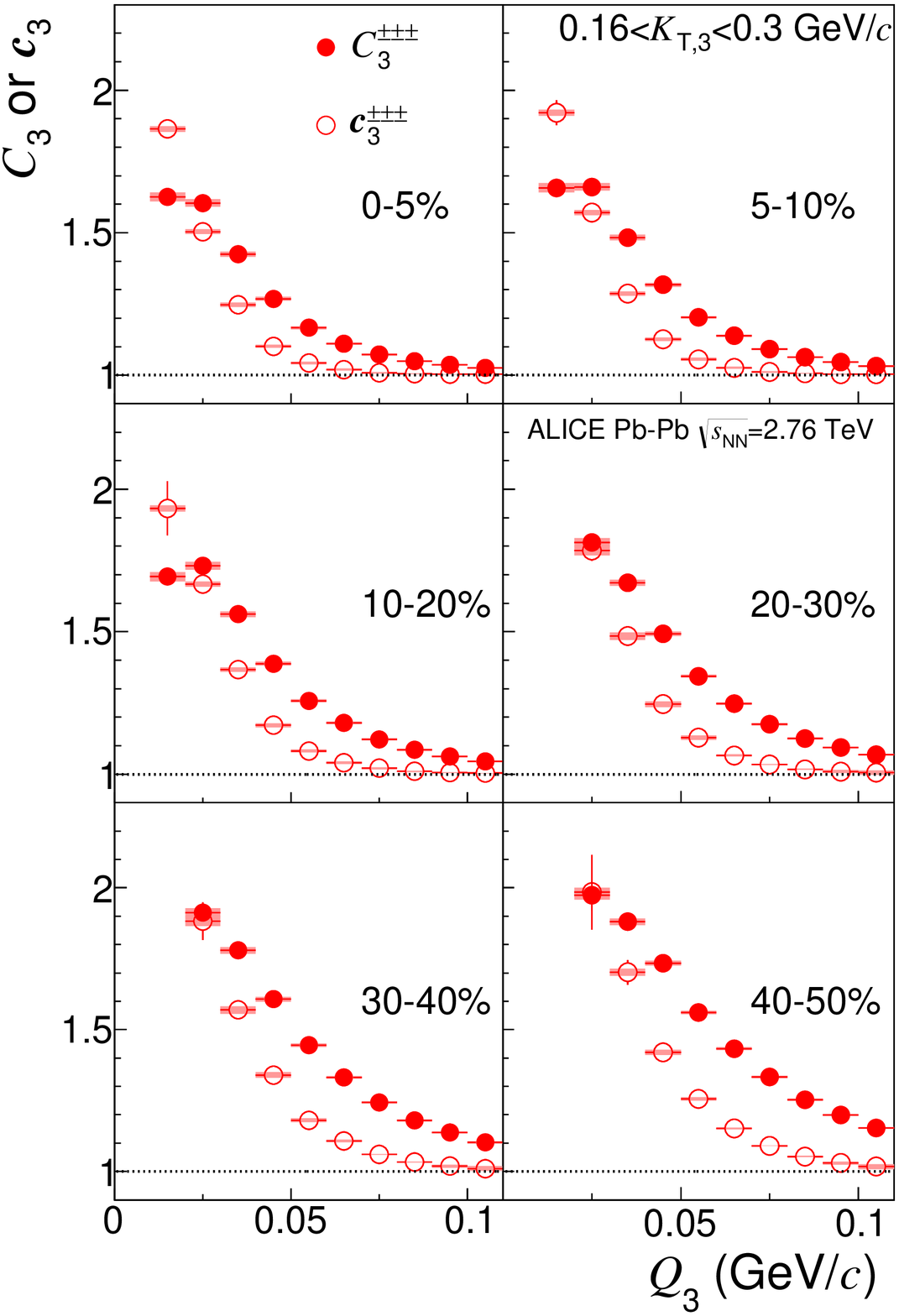}
    \label{fig:C3SC_K1}
  }
  \subfigure[$0.3<K_{\rm T,3}<1.0$ GeV/$c$.]{
    \includegraphics[width=0.49\textwidth]{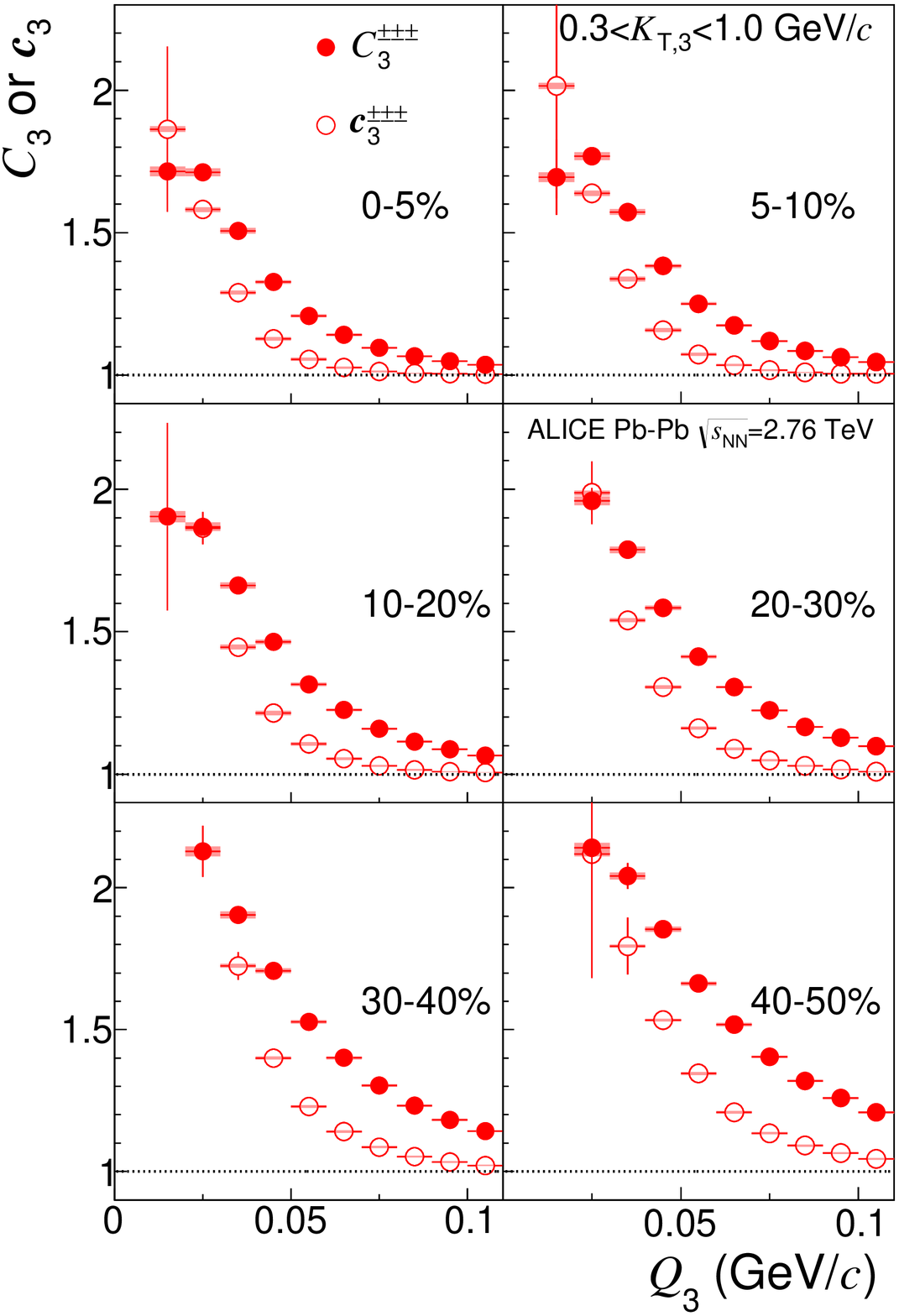}
    \label{fig:C3SC_K2}
  }
  \caption{Same-charge $C_3$ (solid red circles) for each centrality bin for $0.16<K_{\rm T,3}<0.3$ GeV/$c$ (a) and $0.3<K_{\rm T,3}<1.0$ GeV/$c$ (b).  Open points represent the corresponding cumulant correlation functions, ${\rm {\bf c}_3}$.  Shaded bands represent systematic uncertainties.}
\end{figure}

We now present the three-pion same- and mixed-charge correlation functions in two $K_{\rm T,3}=|\bf{p_{\rm T,1}}+\bf{p_{\rm T,2}}+\bf{p_{\rm T,3}}|$/3 bins.  
Two $K_{\rm T,3}$ intervals were chosen such that they divide the number of triplets into two roughly equal halves.
The same-charge three-pion correlations in six centrality bins and two $K_{\rm T,3}$ bins are shown in 
Figs.~\ref{fig:C3SC_K1} and \ref{fig:C3SC_K2}.  
Also shown are the cumulant correlation functions, ${\rm {\bf c}_3}$, for which the two-pion correlations and FSIs are removed.  The 
dilution of correlations caused by $\lambda<1$ is also removed when we consider ${\rm {\bf c}_3}$.
Extraction of the cumulant correlation function, ${\rm {\bf c}_3}$, requires an assumption on the $\lambda$ parameter.  We use the $\lambda$ parameter obtained from two-pion global fits excluding coherence and incorporating an Edgeworth expansion to the full \kt~range ($0<\kt<1.0$).  From central 
to peripheral collisions, $\lambda$ ranges from 0.65 to 0.70.
In Figs.~\ref{fig:C3SC_K1} and \ref{fig:C3SC_K2} we observe that the raw same-charge three-pion correlations are suppressed far below the expected value for fully chaotic emission [$C_3^{\pm\pm\pm}(Q_3=0) < 6$] as 
was similarly seen for $C_2^{\pm\pm}$.  The same-charge cumulant correlation also appears to be suppressed below its 
maximum [${\rm {\bf c}_3}(Q_3=0) < 3$] although a reliable extrapolation to $Q_3=0$ is needed to be sure.  

\begin{figure}
  \subfigure[$0.16<K_{\rm T,3}<0.3$ GeV/$c$.]{
    \includegraphics[width=0.49\textwidth]{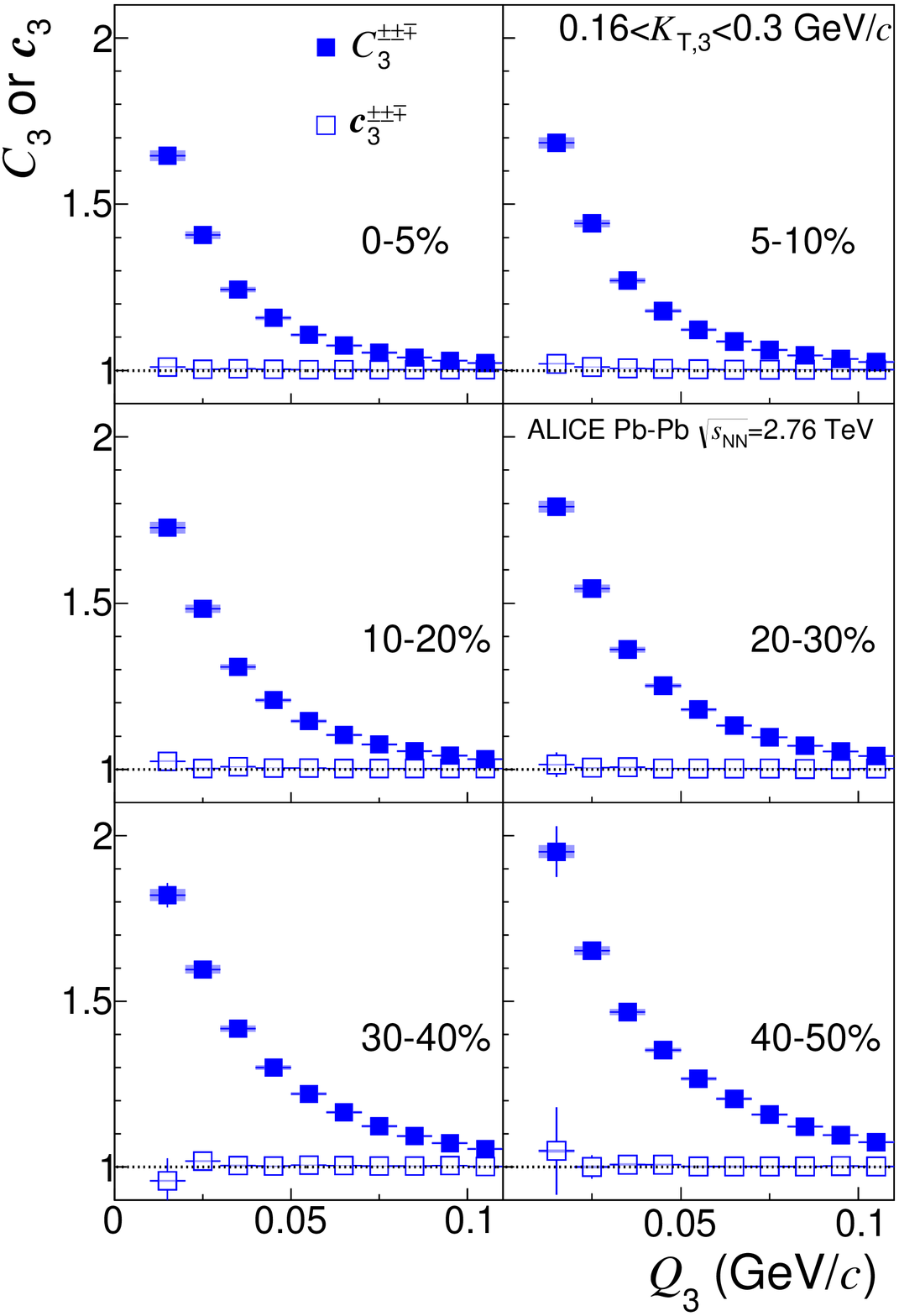}
    \label{fig:C3MC_K1}
  }
  \subfigure[$0.3<K_{\rm T,3}<1.0$ GeV/$c$.]{
    \includegraphics[width=0.49\textwidth]{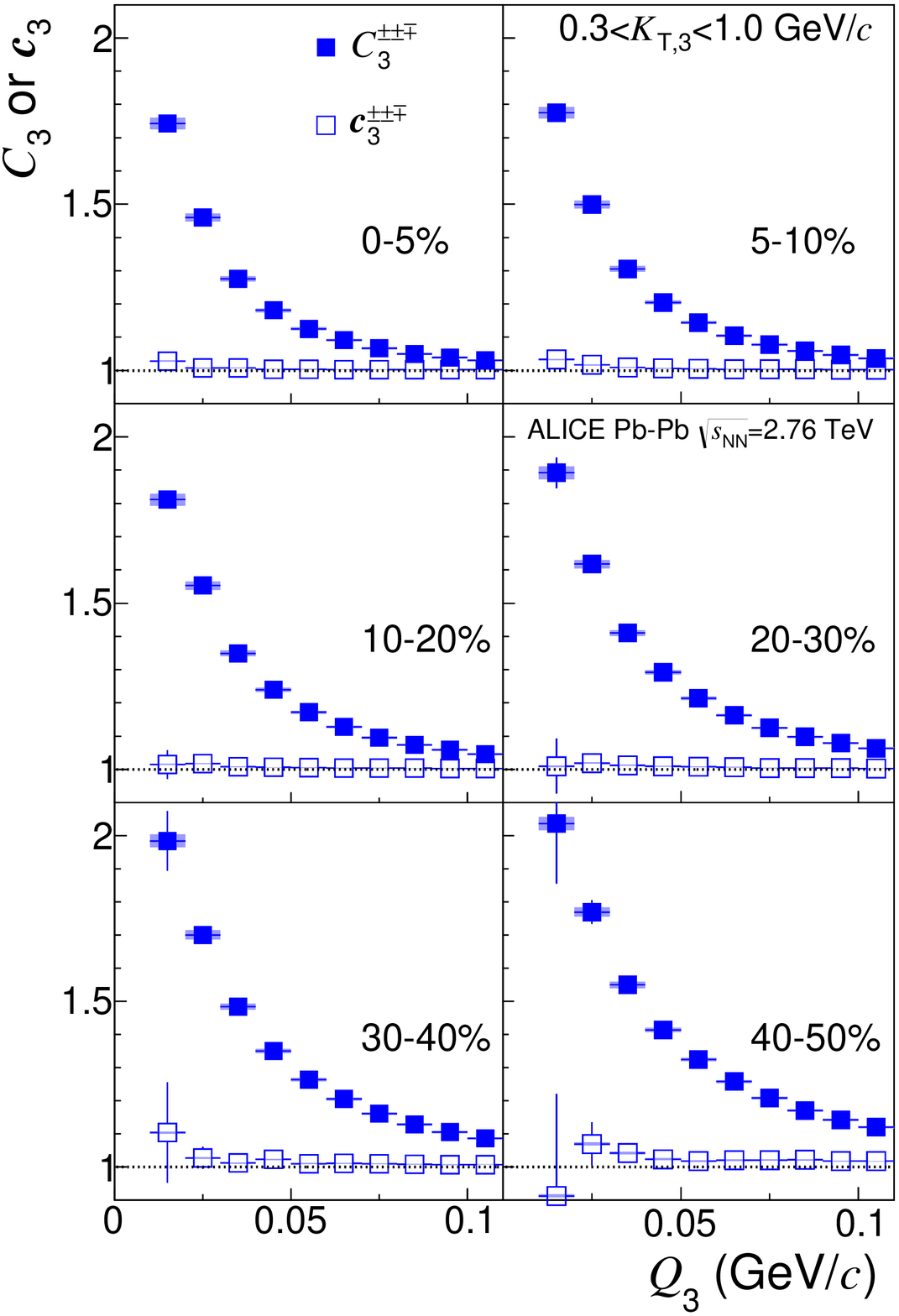}
    \label{fig:C3MC_K2}
  }
  \caption{Mixed-charge $C_3$ (solid blue squares) for each centrality bin for $0.16<K_{\rm T,3}<0.3$ GeV/$c$ (a) and $0.3<K_{\rm T,3}<1.0$ GeV/$c$ (b).  Open squares represent the corresponding cumulant correlation functions, ${\rm {\bf c}_3}$.  Shaded bands represent systematic uncertainties.}
\end{figure}

The mixed-charge three-pion correlations and cumulant correlations in six centrality bins and two $K_{\rm T,3}$ bins are shown in 
Figs.~\ref{fig:C3MC_K1} and \ref{fig:C3MC_K2}.  
For mixed-charge correlations, ${\rm {\bf c}_3}^{\pm\pm\mp}$ is expected to be equal to unity in the presence of only QS and FSIs.  The construction of the cumulant 
correlation function removes FSI effects and the dilution when $\lambda<1$.  The mixed-charge cumulant correlation is largely 
consistent with unity for both $K_{\rm T,3}$ bins although the positive residue for the highest $K_{\rm T,3}$ bin is about 2 times larger than for the lowest bin.  
This demonstrates the validity of asymptotic three-body FSI wave functions for Pb-Pb collisions at the LHC for $Q_3>10$ MeV/$c$.
We note that it may also be possible for a residue to exist for ${\rm {\bf c}_3}^{\pm\pm\mp}$ with charge-constrained coherent states \cite{ALS}.  
The cumulant correlation functions in Figs.~\ref{fig:C3MC_K1} and \ref{fig:C3MC_K2} suggest a residual correlation less than about $1.005$.
The removal of FSI effects is crucial for the interpretation of the intercept of $r_3$.
The successful removal of FSI effects in the mixed-charge three-pion system is demonstrated with the cumulant correlation function in Figs.~\ref{fig:C3MC_K1} and \ref{fig:C3MC_K2}.

The three-pion QS cumulant is compared to the two-pion QS cumulant with $r_3$.  
Unlike fits at the two-particle level alone, the intercept of $r_3$ is more robust to non-Gaussian QS correlations.  By construction, $r_3(Q_3=0)=2.0$ in the 
absence of coherence regardless of the shape of QS correlations \cite{UliChaoticity}. 
To leading order, the relative momentum dependence of $r_3$ was shown to be quartic in the full 
6D approach \cite{UliChaoticity}.  However, owing to limited statistical precision we project $r_3$ onto 1D $Q_3$.

\begin{figure}
  \subfigure[$0.16<K_{\rm T,3}<0.3$ GeV/$c$.]{
    \includegraphics[width=0.49\textwidth]{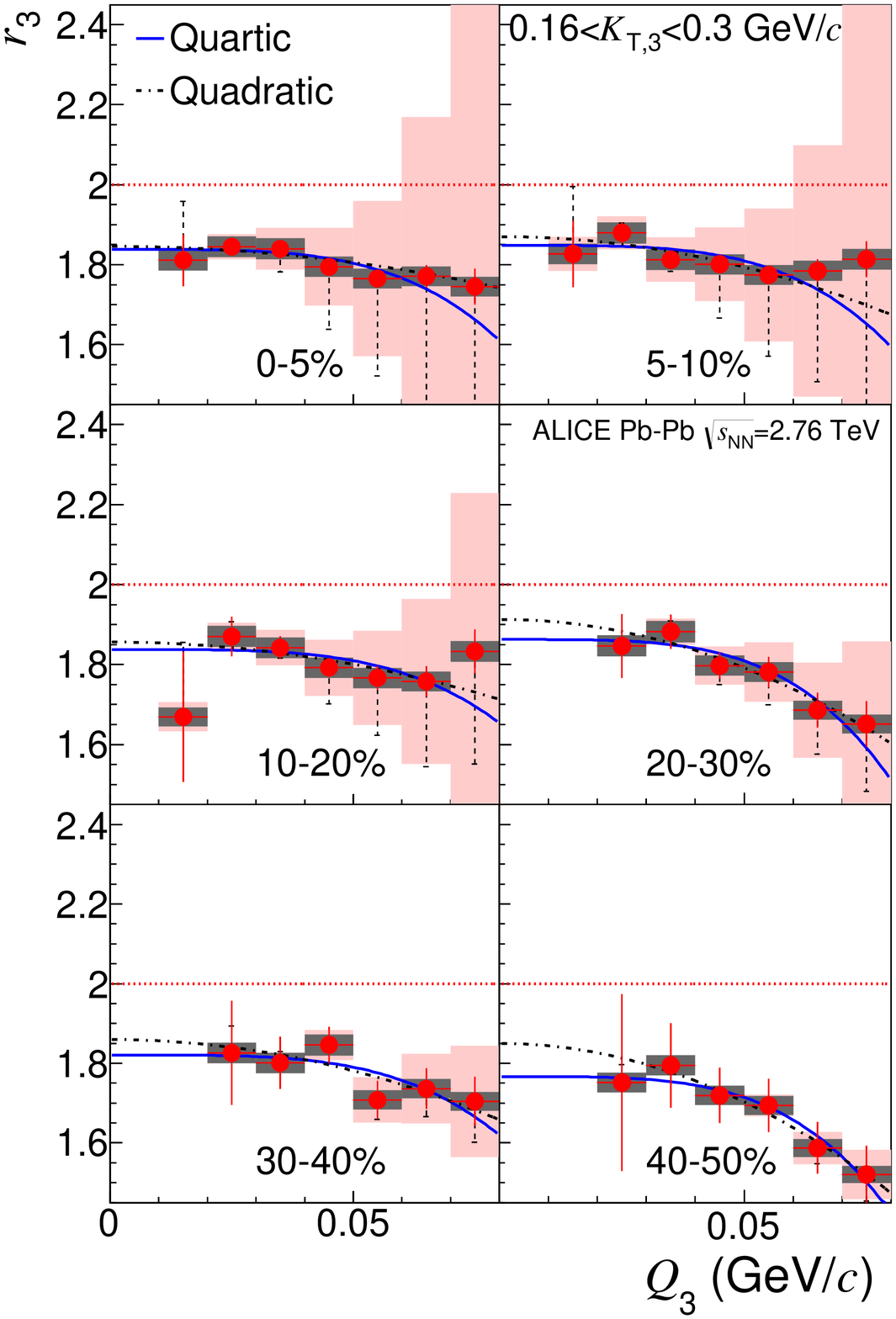}
    \label{fig:r3_Q3_K1}
  }
  \subfigure[$0.3<K_{\rm T,3}<1.0$ GeV/$c$.]{
    \includegraphics[width=0.49\textwidth]{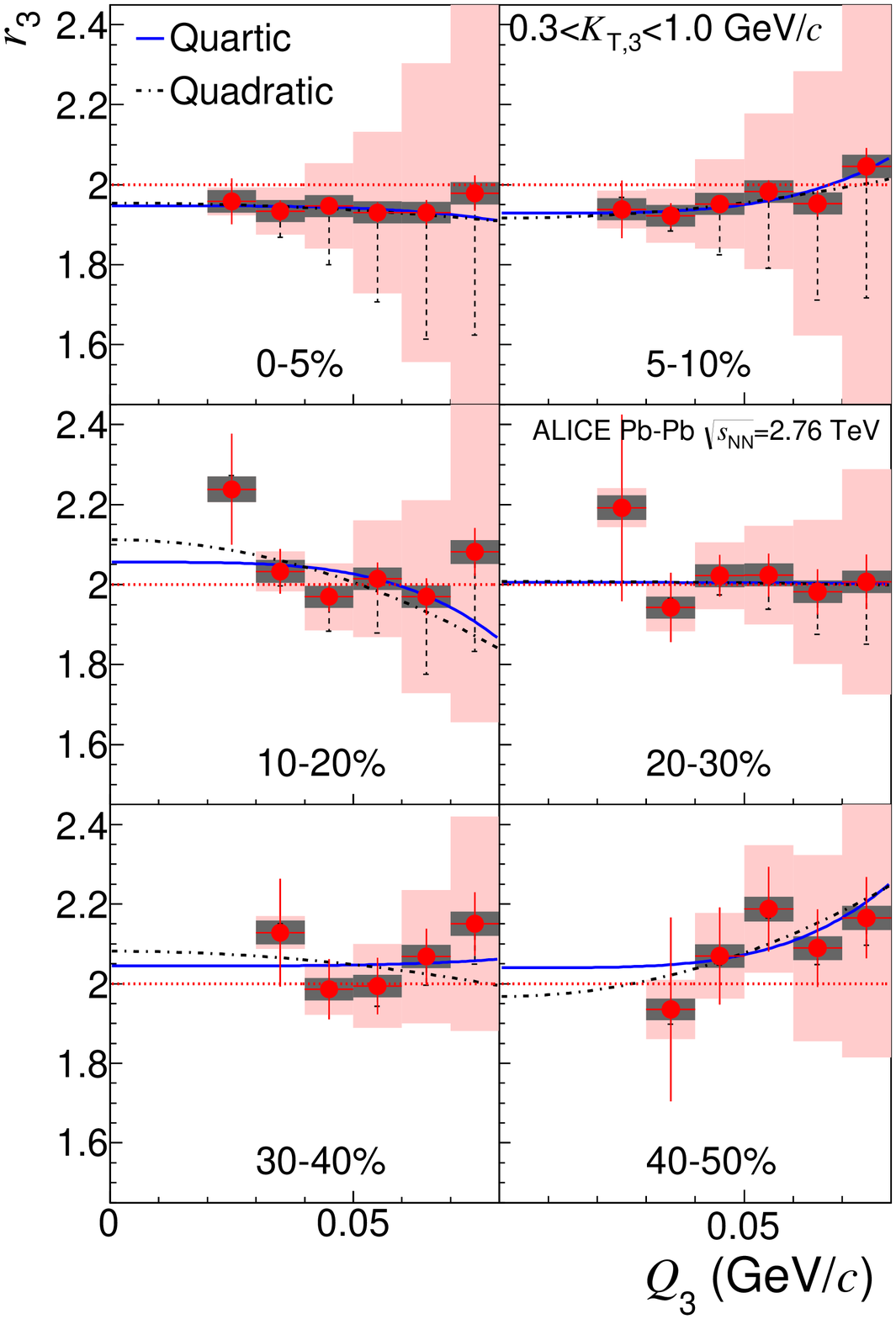}
    \label{fig:r3_Q3_K2}
  }
  \caption{$r_3$ versus $Q_3$ in six centrality bins for $0.16<K_{\rm T,3}<0.3$ GeV/$c$ (a) and $0.3<K_{\rm T,3}<1.0$ GeV/$c$ (b).  
$r_3$ was measured in $5\%$ centrality widths and averaged over the total bin width.  The blue solid line is a quartic fit [Eq.~(\ref{eq:QuarticFit})] and 
the dashed black line is a quadratic fit [Eq.~(\ref{eq:QuadraticFit})].  
The chaotic upper limit [$r_3(Q_3)=2$] is shown 
with the dashed red line.  The shaded gray band represents the systematics owing to PID and momentum resolution.  
The shaded red band represents the uncertainties owing to the choice of $\lambda$ and the residue of the mixed-charge cumulant correlations.  
The dashed line represents uncertainties on the FSI corrections.}
\end{figure}

We now present $r_3$ versus $Q_3$ in Figs.~\ref{fig:r3_Q3_K1} and \ref{fig:r3_Q3_K2} in six centrality bins and two $K_{\rm T,3}$ bins. 
The data are fit with a quartic and quadratic fit as shown by Eqs.~(\ref{eq:QuarticFit}) and (\ref{eq:QuadraticFit}).
The systematic uncertainties at large $Q_3$ are typically larger than $50\%$, while at low $Q_3$ they are much smaller.  
At low $Q_3$, one notices that $r_3$ is further below the chaotic limit (2.0) in 
Fig.~\ref{fig:r3_Q3_K1} than in Fig.~\ref{fig:r3_Q3_K2}.

\begin{figure}
\center
\includegraphics[width=0.49\textwidth]{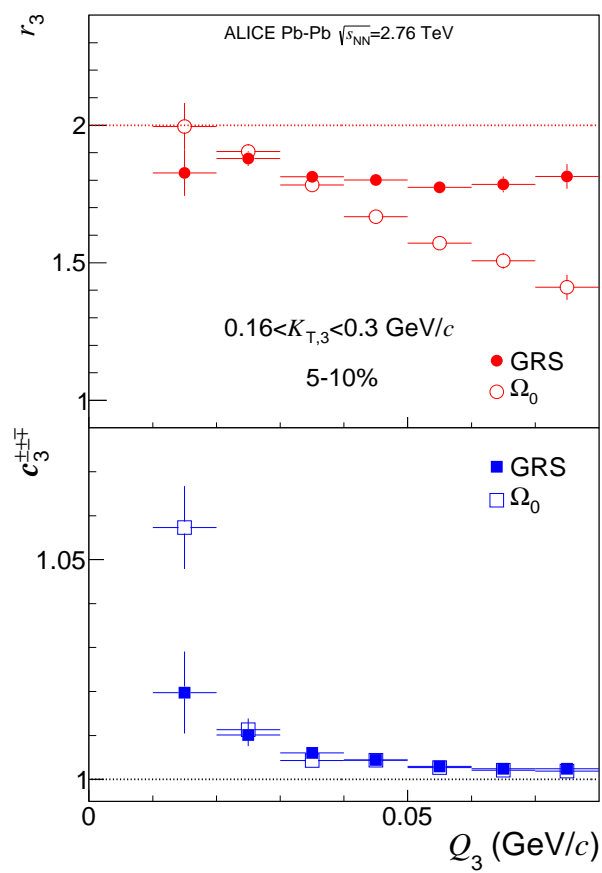}
\caption{In the top panel, $r_3$ versus $Q_3$ is shown with GRS and $\Omega_0$ FSI corrections.  
In the bottom panel, ${\rm {\bf c}_3}^{\pm\pm\mp}$ versus 
$Q_3$ is shown with both FSI corrections.  The centrality and $K_{\rm T,3}$ interval is $5-10\%$ centrality and $0.16<K_{\rm T,3}<0.3$ GeV/$c$, respectively.  Only statistical errors are shown for clarity.}
\label{fig:r3c3_CoulVar}
\end{figure}

The largest systematic uncertainty in Figs.~\ref{fig:r3_Q3_K1} and \ref{fig:r3_Q3_K2} is attributable to the residual correlation of ${\rm {\bf c}_3}^{\pm\pm\mp}$.  
The systematic uncertainties are larger for the higher $K_{\rm T,3}$ bin owing to a larger residual correlation of ${\rm {\bf c}_3}^{\pm\pm\mp}$.
The dashed black lines in Figs.~\ref{fig:r3_Q3_K1} and \ref{fig:r3_Q3_K2} represent the systematic uncertainty owing to FSI corrections.  
It is estimated by the difference in $\Omega_0$ and GRS FSI calculations as was illustrated in Fig.~\ref{fig:K3Omega0GRS}.  Figure \ref{fig:r3c3_CoulVar} compares 
the effect of both FSI corrections on $r_3$ and ${\rm {\bf c}_3}^{\pm\pm\mp}$.  
From the top panel of Fig.~\ref{fig:r3c3_CoulVar} we see that the $\Omega_0$ FSI correction procedure yields an intercept closer to the chaotic 
limit than the GRS procedure.  However, from the bottom panel of Fig.~\ref{fig:r3c3_CoulVar} we see that a large unexplained residual spike remains 
with the $\Omega_0$ FSI correction procedure.  For this reason the GRS procedure was chosen as our standard.  We have also investigated other 
source profile integrations where one obtains larger FSI correlations.  
Such variations, which bring the intercept of $r_3$ to the chaotic limit, simultaneously cause a large overcorrection of the mixed-charge three-pion cumulant, 
${\rm {\bf c}_3}^{\pm\pm\mp}(Q_3\sim0) \sim 0.96$.

\begin{figure}
\center
\includegraphics[width=0.49\textwidth]{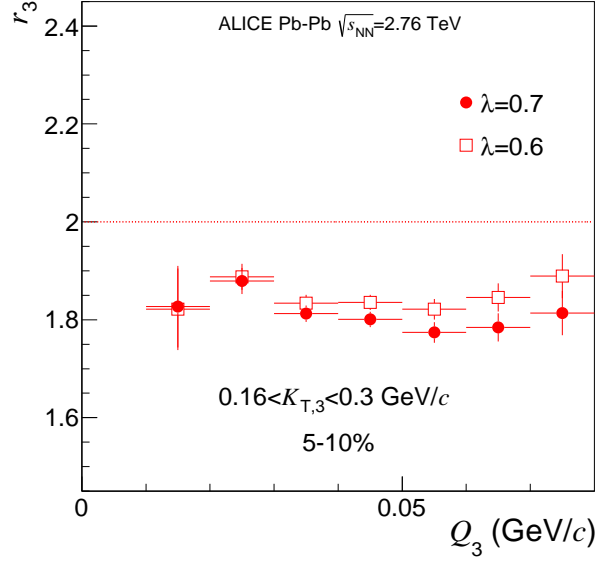}
\caption{$r_3$ versus $Q_3$ is shown with two different assumptions on the $\lambda$ parameter for 
$5-10\%$ centrality and for $0.16<K_{\rm T,3}<0.3$ GeV/$c$.  Only statistical errors are shown for clarity.}
\label{fig:Fig_r3_LamVar}
\end{figure}

In Fig.~\ref{fig:Fig_r3_LamVar} we show $r_3$ with two different assumptions on the $\lambda$ parameter.
The default value of 0.7 is compared to 0.6 in Fig.~\ref{fig:Fig_r3_LamVar}.  The default value was motivated by Edgeworth fits at the two-pion level 
as was shown in Fig.~\ref{fig:C2fitParams}.
The effect of the chosen $\lambda$ parameter only has non-negligible effect at large $Q_3$ and in central collisions where 
the cumulant correlation is small, ${\rm {\bf c}_3}^{\pm\pm\pm} \sim 1.0$.

We see that the $Q_3$ dependence of $r_3$ is largely uncertain for the more central collisions.  This is caused by the uncertainty in 
isolating the three-pion QS cumulant when the cumulant correlation itself is small, ${\rm {\bf c}_3}^{\pm\pm\pm} \sim 1.0$.  
A quartic [Eq.~(\ref{eq:QuarticFit})] and quadratic [Eq.~(\ref{eq:QuadraticFit})] fit are shown in Figs.~\ref{fig:r3_Q3_K1} 
and \ref{fig:r3_Q3_K2} and are summarized in Tables \ref{tab:r3_Q3_Quartic} and \ref{tab:r3_Q3_Quadratic}, 
respectively.
\begin{table}
\centering
	\begin{tabular}{| c | c | c |}
		\hline
              Low $K_{\rm T,3}$ & $I$ $\pm$ stat $\pm$ syst & $a\times10^{3}$ (GeV/$c$)$^{-4}$ \\ \hline
$0-5\%$ & $1.84 \pm 0.01 \pm 0.03$ & $3.0 \pm 0.6 \pm 16.4$ \\ \hline
$5-10\%$ & $1.85 \pm 0.01 \pm 0.05$ & $3.4 \pm 0.7 \pm 13.0$ \\ \hline
$10-20\%$ & $1.84 \pm 0.02 \pm 0.03$ & $2.4 \pm 0.9 \pm 8.1$ \\ \hline
$20-30\%$ & $1.86 \pm 0.03 \pm 0.01$ & $4.6 \pm 1.0 \pm 3.7$ \\ \hline
$30-40\%$ & $1.82 \pm 0.04 \pm 0.03$ & $2.7 \pm 1.3 \pm 2.8$ \\ \hline
$40-50\%$ & $1.77 \pm 0.05 \pm 0.01$ & $4.8 \pm 1.6 \pm 1.1$ \\ \hline
$0-50\%$ & $1.83 \pm 0.01 \pm 0.03$ & $3.5 \pm 0.4 \pm 7.5$ \\ \hline
              High $K_{\rm T,3}$ &  &  \\ \hline
$0-5\%$ & $1.95 \pm 0.02 \pm 0.02$ & $0.5 \pm 0.7 \pm 10.1$ \\ \hline
$5-10\%$ & $1.93 \pm 0.02 \pm 0.01$ & $-1.8 \pm 0.8 \pm 8.4$ \\ \hline
$10-20\%$ & $2.06 \pm 0.03 \pm 0.07$ & $2.3 \pm 1.1 \pm 5.7$ \\ \hline
$20-30\%$ & $2.01 \pm 0.04 \pm 0.01$ & $0.0 \pm 1.3 \pm 3.0$ \\ \hline
$30-40\%$ & $2.04 \pm 0.06 \pm 0.05$ & $-0.2 \pm 1.8 \pm 3.0$ \\ \hline
$40-50\%$ & $2.04 \pm 0.09 \pm 0.04$ & $-2.6 \pm 2.4 \pm 1.4$ \\ \hline
$0-50\%$ & $2.00 \pm 0.02 \pm 0.03$ & $-0.3 \pm 0.6 \pm 5.3$ \\ \hline
        \end{tabular}	
	\caption{Quartic $r_3$ fit parameters from Figs.~\ref{fig:r3_Q3_K1} and \ref{fig:r3_Q3_K2}.  The centrality averaged values 
are also shown.  Statistical and systematic uncertainties are shown.  Low $K_{\rm T,3}$ refers to $0.16<K_{\rm T,3}<0.3$ GeV/$c$.  
High $K_{\rm T,3}$ refers to $0.3<K_{\rm T,3}<1.0$ GeV/$c$.}
        \label{tab:r3_Q3_Quartic}
\end{table}

\begin{table}
\centering
	\begin{tabular}{| c | c | c |}
		\hline
                Low $K_{\rm T,3}$ & $I$ $\pm$ stat $\pm$ syst & $a\times10^{1}$ (GeV/$c$)$^{-2}$ \\ \hline
$0-5\%$ & $1.85 \pm 0.02 \pm 0.11$ & $0.9 \pm 0.3 \pm 6.7$ \\ \hline
$5-10\%$ & $1.87 \pm 0.02 \pm 0.12$ & $1.6 \pm 0.4 \pm 5.8$ \\ \hline
$10-20\%$ & $1.86 \pm 0.03 \pm 0.09$ & $1.2 \pm 0.5 \pm 4.1$ \\ \hline
$20-30\%$ & $1.91 \pm 0.04 \pm 0.04$ & $2.5 \pm 0.6 \pm 1.9$ \\ \hline
$30-40\%$ & $1.86 \pm 0.05 \pm 0.07$ & $1.7 \pm 0.8 \pm 1.7$ \\ \hline
$40-50\%$ & $1.85 \pm 0.08 \pm 0.01$ & $3.2 \pm 1.0 \pm 0.7$ \\ \hline
$0-50\%$ & $1.87 \pm 0.02 \pm 0.07$ & $1.8 \pm 0.3 \pm 3.5$ \\ \hline
                High $K_{\rm T,3}$ &  &  \\ \hline
$0-5\%$ & $1.95 \pm 0.03 \pm 0.06$ & $0.4 \pm 0.5 \pm 4.8$ \\ \hline
$5-10\%$ & $1.92 \pm 0.03 \pm 0.07$ & $-0.8 \pm 0.5 \pm 4.4$ \\ \hline
$10-20\%$ & $2.11 \pm 0.05 \pm 0.12$ & $2.0 \pm 0.7 \pm 3.5$ \\ \hline
$20-30\%$ & $2.01 \pm 0.07 \pm 0.05$ & $0.1 \pm 0.9 \pm 1.9$ \\ \hline
$30-40\%$ & $2.08 \pm 0.09 \pm 0.13$ & $0.6 \pm 1.3 \pm 2.4$ \\ \hline
$40-50\%$ & $1.97 \pm 0.15 \pm 0.07$ & $-2.2 \pm 1.9 \pm 1.1$ \\ \hline
$0-50\%$ & $2.01 \pm 0.03 \pm 0.08$ & $0.0 \pm 0.5 \pm 3.0$ \\ \hline
        \end{tabular}	
	\caption{Quadratic $r_3$ fit parameters from Figs.~\ref{fig:r3_Q3_K1} and \ref{fig:r3_Q3_K2}.  The centrality averaged values 
are also shown.  Statistical and systematic uncertainties are shown.  Low $K_{\rm T,3}$ refers to $0.16<K_{\rm T,3}<0.3$ GeV/$c$.  
High $K_{\rm T,3}$ refers to $0.3<K_{\rm T,3}<1.0$ GeV/$c$.}
        \label{tab:r3_Q3_Quadratic}
\end{table}
Given the large uncertainties at large $Q_3$, $r_3$ does not change significantly with centrality and is equally well described 
by quartic and quadratic fits.  The centrality averaged fit values are 
also given in Tables \ref{tab:r3_Q3_Quartic} and \ref{tab:r3_Q3_Quadratic}.  

From the intercepts of $r_{3}$ at $Q_3=0$ presented in Tables \ref{tab:r3_Q3_Quartic} and \ref{tab:r3_Q3_Quadratic}, 
the corresponding coherent fractions ($G$) may be extracted using Eq.~(\ref{eq:Gfromr3}).  For low $K_{\rm T,3}$, 
the centrality averaged intercepts ($0-50\%$) of 
$r_3$ may correspond to coherent fractions of $28\% \pm 3\%$ and $24\% \pm 9\%$ for quartic and quadratic intercepts, respectively.
For high $K_{\rm T,3}$, the corresponding coherent fractions are consistent with zero for both quartic and quadratic fits.  Given the systematic uncertainties at large $Q_3$, both quartic and 
quadratic fits provide a good description of $r_3$.  We estimate the average coherent fraction at low $K_{\rm T,3}$ using both quartic and quadratic fits as well as their uncertainties as: $(G^{\rm quartic}+\delta G^{\rm quartic} + G^{\rm quadratic}-\delta G^{\rm quadratic})/2$.  The average coherent fraction at low 
$K_{\rm T,3}$ is estimated to be $23\% \pm 8\%$.

As a sanity check, we also reconstructed $r_3$ in \textsc{hijing} including the simulated response of the ALICE detector.  
\textsc{hijing} does not contain QS nor FSIs.  
We used a known symmetric and fully chaotic QS+FSI correlation as a pair/triplet fill weight.  
The same code developed for this analysis was used in this procedure.  The reconstructed $r_3$ for both $K_{\rm T,3}$ bins 
was consistent with the chaotic limit for all $Q_3$.

\section{Conclusions}
Two- and three-pion quantum statistical correlations in Pb-Pb collisions at $\sqrt{s_{NN}}=2.76$ TeV have been presented.
Same-charge as well as mixed-charge combinations were shown for both two- and three-pion correlations.  While same-charge 
correlations uniquely display the effect of quantum interference, mixed-charge correlations provide an important constraint 
on the $\lambda$ parameter and FSI corrections in this analysis.  

At the two-pion level, we find that while same-charge correlations change rapidly with \kt, mixed-charge correlations change 
very little.  A comparison of mixed-charge correlations to \textsc{therminator} suggests that the $\lambda$ parameter changes 
very little with \kt.  Global fits to same- and mixed-charge correlations at the 
two-pion level alone are inconclusive in determining the presence of coherence owing to the unknown non-Gaussian features of the same-charge correlation function. 

Three-pion mixed-charge correlations are very well described by the combination of QS and FSI correlations.  While the mixed-charge 
three-pion cumulant correlation is largely consistent with unity, the same-charge three-pion cumulant shows a significant QS correlation.

The comparison of the three-pion cumulant to the two-pion cumulant is measured with $r_3$.  Unlike fits at the two-pion level alone, 
the intercept of $r_3$ is more robust to non-Gaussian Bose-Einstein correlations.  We find a clear suppression of $r_3$ below the 
chaotic limit for low $K_{\rm T,3}$ while being much more consistent with the chaotic limit for high $K_{\rm T,3}$.  Incomplete FSI removal, momentum 
resolution correction, and pion misidentification can also cause an apparent suppression of $r_3$.  However, the $K_{\rm T,3}$ dependencies of 
the $r_3$ intercepts go in the opposite direction than would be expected from such effects.

Given the large uncertainties at large $Q_3$, $r_3$ does not change significantly with centrality.  For low triplet momentum, 
the centrality averaged intercepts of $r_3$ may correspond to a coherent fraction of $23\% \pm 8\%$.  For high triplet momentum the 
intercepts of $r_3$ yield a coherent fraction consistent with zero.

The suppression of three-pion as compared to two-pion Bose-Einstein correlations as measured by $r_3$ seems to suggest a 
finite coherent component to pion production in heavy-ion collisions.  It 
is significant at low triplet momentum while vanishing for high triplet momentum.  This observation 
is qualitatively consistent with the formation of a Bose-Einstein condensate which is expected to radiate coherently at low momentum.
More experimental and theoretical work is needed to rule out alternative explanations.  Other measurements such 
as the single-pion spectra should provide additional information on this subject.
We also note that the ALICE 
single-pion spectra indicate a small excess of pion production as compared to several hydrodynamic calculations for $\pt<0.4$ GeV/$c$ 
\cite{ALICEpionspectra}.  The mean \pt~of pions for low $Q_3$ in our lowest and highest $K_{\rm T,3}$ bin is about 0.24 and 0.38 GeV/$c$, respectively.  
The excess in the single-pion spectra may be related to the coherent fractions extracted in this analysis.

\newenvironment{acknowledgement}{\relax}{\relax}
\begin{acknowledgement}
\section*{Acknowledgments}
We would like to thank Richard Lednick{\' y}, Ulrich Heinz, Tam{\' a}s Cs{\" o}rg{\H o}, M{\' a}t{\' e} Csan{\' a}d, and Yuri Sinyukov for numerous helpful discussions.

The ALICE collaboration would like to thank all its engineers and technicians for their invaluable contributions to the construction of the experiment and the CERN accelerator teams for the outstanding performance of the LHC complex.
\\
The ALICE collaboration gratefully acknowledges the resources and support provided by all Grid centres and the Worldwide LHC Computing Grid (WLCG) collaboration.
\\
The ALICE collaboration acknowledges the following funding agencies for their support in building and
running the ALICE detector:
 \\
State Committee of Science,  World Federation of Scientists (WFS)
and Swiss Fonds Kidagan, Armenia,
 \\
Conselho Nacional de Desenvolvimento Cient\'{\i}fico e Tecnol\'{o}gico (CNPq), Financiadora de Estudos e Projetos (FINEP),
Funda\c{c}\~{a}o de Amparo \`{a} Pesquisa do Estado de S\~{a}o Paulo (FAPESP);
 \\
National Natural Science Foundation of China (NSFC), the Chinese Ministry of Education (CMOE)
and the Ministry of Science and Technology of China (MSTC);
 \\
Ministry of Education and Youth of the Czech Republic;
 \\
Danish Natural Science Research Council, the Carlsberg Foundation and the Danish National Research Foundation;
 \\
The European Research Council under the European Community's Seventh Framework Programme;
 \\
Helsinki Institute of Physics and the Academy of Finland;
 \\
French Grant No.~CNRS-IN2P3, the ``Region Pays de Loire,'' ``Region Alsace,'' ``Region Auvergne,'' and CEA, France;
 \\
German BMBF and the Helmholtz Association;
\\
General Secretariat for Research and Technology, Ministry of
Development, Greece;
\\
Hungarian OTKA and National Office for Research and Technology (NKTH);
 \\
Department of Atomic Energy and Department of Science and Technology of the Government of India;
 \\
Istituto Nazionale di Fisica Nucleare (INFN) and Centro Fermi -
Museo Storico della Fisica e Centro Studi e Ricerche ``Enrico Fermi,'' Italy;
 \\
MEXT Grant-in-Aid for Specially Promoted Research, Ja\-pan;
 \\
Joint Institute for Nuclear Research, Dubna;
 \\
National Research Foundation of Korea (NRF);
 \\
CONACYT, DGAPA, M\'{e}xico, ALFA-EC and the EPLANET Program
(European Particle Physics Latin American Network)
 \\
Stichting voor Fundamenteel Onderzoek der Materie (FOM) and the Nederlandse Organisatie voor Wetenschappelijk Onderzoek (NWO), Netherlands;
 \\
Research Council of Norway (NFR);
 \\
Polish National Science Centre, Grant No.~2012/05/N/ST2/02757,2013/08/M/ST2/00598;
 \\
National Science Centre, Poland;
 \\
 Ministry of National Education/Institute for Atomic Physics and CNCS-UEFISCDI, Romania;
 \\
Ministry of Education and Science of Russian Federation, Russian
Academy of Sciences, Russian Federal Agency of Atomic Energy,
Russian Federal Agency for Science and Innovations, and The Russian
Foundation for Basic Research;
 \\
Ministry of Education of Slovakia;
 \\
Department of Science and Technology, South Africa;
 \\
CIEMAT, EELA, Ministerio de Econom\'{i}a y Competitividad (MINECO) of Spain; Xunta de Galicia (Conseller\'{\i}a de Educaci\'{o}n),
CEA\-DEN, Cubaenerg\'{\i}a, Cuba; IAEA (International Atomic Energy Agency);
 \\
Swedish Research Council (VR) and Knut $\&$ Alice Wallenberg
Foundation (KAW);
 \\
Ukraine Ministry of Education and Science;
 \\
United Kingdom Science and Technology Facilities Council (STFC);
 \\
The United States Department of Energy, the United States National
Science Foundation, the State of Texas, and the State of Ohio.

\end{acknowledgement}


\newpage
\appendix
\section{Relation of $N_3$ to $N_3^{QS}$}
The measurement of the true three-pion correlation is more involved when the ``dilution'' parameter, $\lambda$, is less than unity. 
In the core/halo picture \cite{CoreHalo}, the effective intercept parameter is given by $\lambda_*$ which represents the fraction of pairs interacting at low relative momentum via QS+FSIs above the resolvable threshold $q_{\rm min}$.  
Here, $\lambda$ includes the additional dilution caused by secondary contamination and pion mis-identification.
The probability of choosing $N$ particles from the interacting core is $\lambda^{N/2}$.  
In general, $\lambda$ is less than unity.  This means that despite measuring three-pions from the same event, there will be a 
fraction of triplets which do not represent a true three-particle interaction.  These feed-up contributions must be removed.
In general, the measured three-particle distribution will take on the form
\begin{eqnarray}
N_3(p_1,p_2,p_3) &=& f_1'N_1(p_1)N_1(p_2)N_1(p_3) \nonumber \\
&+& f_2'\big[N_2^{\rm true}(p_1,p_2)N_1(p_3) + N_2^{\rm true}(p_3,p_1)N_1(p_2) + N_2^{\rm true}(p_2,p_3)N_1(p_1) \big] \nonumber \\
&+& f_3'N_3^{\rm true}(p_1,p_2,p_3),
\end{eqnarray}
where $f_1',f_2',f_3'$ represent the fraction of triplets for which none interact, two interact, and all three interact, respectively.
The probability that all three are from the noninteracting halo is then $(1-\lambda^{1/2})^3$.  The probability that only 
one is from the interacting core is $3\lambda^{1/2}(1-\lambda^{1/2})^2$.  Therefore, 
\begin{equation}
f_1' = (1-\lambda^{1/2})^3 + 3\lambda^{1/2}(1-\lambda^{1/2})^2.
\end{equation}
The probability that two are from the interacting core is $\lambda(1-\lambda^{1/2})$.  
Therefore,
\begin{equation}
f_2' = \lambda(1-\lambda^{1/2}).
\end{equation}
Finally, the probability that all three are from the interacting core is $\lambda^{3/2}$. Therefore,
\begin{equation}
f_3' = \lambda^{3/2}
\end{equation}
Now we can write the equation expressing the triplet distribution in terms of the true distributions:
\begin{eqnarray}
N_3(p_1,p_2,p_3) &=& [(1-\lambda^{1/2})^3 + 3\lambda^{1/2}(1-\lambda^{1/2})^2]N_1(p_1)N_1(p_2)N_1(p_3)\nonumber \\ 
&+& \lambda(1-\lambda^{1/2})\big[N_2^{\rm true}(p_1,p_2)N_1(p_3) + N_2^{\rm true}(p_3,p_1)N_1(p_2) + N_2^{\rm true}(p_2,p_3)N_1(p_1)\big] \nonumber \\
&+& \lambda^{3/2}N_3^{\rm true}(p_1,p_2,p_3).
\label{eq:IntermediateStep}
\end{eqnarray}
$N_2^{\rm true}$ is related to the measured $N_2$ through Eq.~(\ref{eq:N2QS}) with ${\mathcal N}=1$.
Finally, we assume a factorization of the three-pion FSI correlation, $K_3$, from the QS correlation.
We can now form a relation between the QS three-pion distribution and the measured distributions:
\begin{eqnarray}
N_3(p_1,p_2,p_3) &=& [(1-\lambda^{1/2})^3 + 3\lambda^{1/2}(1-\lambda^{1/2})^2 - 3(1-\lambda^{1/2})(1-\lambda)]N_1(p_1)N_1(p_2)N_1(p_3) \nonumber \\ 
&+& (1-\lambda^{1/2})\big[N_2(p_1,p_2)N_1(p_3) + N_2(p_3,p_1)N_1(p_2) + N_2(p_2,p_3)N_1(p_1)\big] \nonumber \\
&+& \lambda^{3/2}K_3N_3^{QS}(p_1,p_2,p_3).
\end{eqnarray}

\newpage
\section{The ALICE Collaboration}
\label{app:collab}



\begingroup
\small
\begin{flushleft}
B.~Abelev\Irefn{org74}\And
J.~Adam\Irefn{org38}\And
D.~Adamov\'{a}\Irefn{org82}\And
M.M.~Aggarwal\Irefn{org86}\And
G.~Aglieri~Rinella\Irefn{org34}\And
M.~Agnello\Irefn{org92}\textsuperscript{,}\Irefn{org109}\And
A.G.~Agocs\Irefn{org132}\And
A.~Agostinelli\Irefn{org26}\And
N.~Agrawal\Irefn{org45}\And
Z.~Ahammed\Irefn{org128}
% ...
\renewcommand\labelenumi{\textsuperscript{\theenumi}~}

\section*{Affiliation notes}
\renewcommand\theenumi{\roman{enumi}}
\begin{Authlist}
\item \Adef{0}Deceased
\item \Adef{idp1135680}{Also at: St-Petersburg State Polytechnical University}
\item \Adef{idp3050192}{Also at: Department of Applied Physics, Aligarh Muslim University, Aligarh, India}
\item \Adef{idp3756272}{Also at: M.V. Lomonosov Moscow State University, D.V. Skobeltsyn Institute of Nuclear Physics, Moscow, Russia}
\item \Adef{idp4007808}{Also at: University of Belgrade, Faculty of Physics and "Vin\v{c}a" Institute of Nuclear Sciences, Belgrade, Serbia}
\item \Adef{idp4828624}{Also at: Institute of Theoretical Physics, University of Wroclaw, Wroclaw, Poland}
\item \Adef{idp5736048}{Also at: the University of Kansas, Lawrence, KS, United States}
\end{Authlist}

\section*{Collaboration Institutes}
\renewcommand\theenumi{\arabic{enumi}~}
\begin{Authlist}

\item \Idef{org1}A.I. Alikhanyan National Science Laboratory (Yerevan Physics Institute) Foundation, Yerevan, Armenia
\item \Idef{org2}Benem\'{e}rita Universidad Aut\'{o}noma de Puebla, Puebla, Mexico
\item \Idef{org3}Bogolyubov Institute for Theoretical Physics, Kiev, Ukraine
\item \Idef{org4}Bose Institute, Department of Physics and Centre for Astroparticle Physics and Space Science (CAPSS), Kolkata, India
\item \Idef{org5}Budker Institute for Nuclear Physics, Novosibirsk, Russia
\item \Idef{org6}California Polytechnic State University, San Luis Obispo, CA, United States
\item \Idef{org7}Central China Normal University, Wuhan, China
\item \Idef{org8}Centre de Calcul de l'IN2P3, Villeurbanne, France
\item \Idef{org9}Centro de Aplicaciones Tecnol\'{o}gicas y Desarrollo Nuclear (CEADEN), Havana, Cuba
\item \Idef{org10}Centro de Investigaciones Energ\'{e}ticas Medioambientales y Tecnol\'{o}gicas (CIEMAT), Madrid, Spain
\item \Idef{org11}Centro de Investigaci\'{o}n y de Estudios Avanzados (CINVESTAV), Mexico City and M\'{e}rida, Mexico
\item \Idef{org12}Centro Fermi - Museo Storico della Fisica e Centro Studi e Ricerche ``Enrico Fermi'', Rome, Italy
\item \Idef{org13}Chicago State University, Chicago, USA
\item \Idef{org14}Commissariat \`{a} l'Energie Atomique, IRFU, Saclay, France
\item \Idef{org15}COMSATS Institute of Information Technology (CIIT), Islamabad, Pakistan
\item \Idef{org16}Departamento de F\'{\i}sica de Part\'{\i}culas and IGFAE, Universidad de Santiago de Compostela, Santiago de Compostela, Spain
\item \Idef{org17}Department of Physics and Technology, University of Bergen, Bergen, Norway
\item \Idef{org18}Department of Physics, Aligarh Muslim University, Aligarh, India
\item \Idef{org19}Department of Physics, Ohio State University, Columbus, OH, United States
\item \Idef{org20}Department of Physics, Sejong University, Seoul, South Korea
\item \Idef{org21}Department of Physics, University of Oslo, Oslo, Norway
\item \Idef{org22}Dipartimento di Fisica dell'Universit\`{a} 'La Sapienza' and Sezione INFN Rome
\item \Idef{org23}Dipartimento di Fisica dell'Universit\`{a} and Sezione INFN, Cagliari, Italy
\item \Idef{org24}Dipartimento di Fisica dell'Universit\`{a} and Sezione INFN, Trieste, Italy
\item \Idef{org25}Dipartimento di Fisica dell'Universit\`{a} and Sezione INFN, Turin, Italy
\item \Idef{org26}Dipartimento di Fisica e Astronomia dell'Universit\`{a} and Sezione INFN, Bologna, Italy
\item \Idef{org27}Dipartimento di Fisica e Astronomia dell'Universit\`{a} and Sezione INFN, Catania, Italy
\item \Idef{org28}Dipartimento di Fisica e Astronomia dell'Universit\`{a} and Sezione INFN, Padova, Italy
\item \Idef{org29}Dipartimento di Fisica `E.R.~Caianiello' dell'Universit\`{a} and Gruppo Collegato INFN, Salerno, Italy
\item \Idef{org30}Dipartimento di Scienze e Innovazione Tecnologica dell'Universit\`{a} del  Piemonte Orientale and Gruppo Collegato INFN, Alessandria, Italy
\item \Idef{org31}Dipartimento Interateneo di Fisica `M.~Merlin' and Sezione INFN, Bari, Italy
\item \Idef{org32}Division of Experimental High Energy Physics, University of Lund, Lund, Sweden
\item \Idef{org33}Eberhard Karls Universit\"{a}t T\"{u}bingen, T\"{u}bingen, Germany
\item \Idef{org34}European Organization for Nuclear Research (CERN), Geneva, Switzerland
\item \Idef{org35}Excellence Cluster Universe, Technische Universit\"{a}t M\"{u}nchen, Munich, Germany
\item \Idef{org36}Faculty of Engineering, Bergen University College, Bergen, Norway
\item \Idef{org37}Faculty of Mathematics, Physics and Informatics, Comenius University, Bratislava, Slovakia
\item \Idef{org38}Faculty of Nuclear Sciences and Physical Engineering, Czech Technical University in Prague, Prague, Czech Republic
\item \Idef{org39}Faculty of Science, P.J.~\v{S}af\'{a}rik University, Ko\v{s}ice, Slovakia
\item \Idef{org40}Frankfurt Institute for Advanced Studies, Johann Wolfgang Goethe-Universit\"{a}t Frankfurt, Frankfurt, Germany
\item \Idef{org41}Gangneung-Wonju National University, Gangneung, South Korea
\item \Idef{org42}Gauhati University, Department of Physics, Guwahati, India
\item \Idef{org43}Helsinki Institute of Physics (HIP), Helsinki, Finland
\item \Idef{org44}Hiroshima University, Hiroshima, Japan
\item \Idef{org45}Indian Institute of Technology Bombay (IIT), Mumbai, India
\item \Idef{org46}Indian Institute of Technology Indore, Indore (IITI), India
\item \Idef{org47}Inha University, College of Natural Sciences
\item \Idef{org48}Institut de Physique Nucleaire d'Orsay (IPNO), Universite Paris-Sud, CNRS-IN2P3, Orsay, France
\item \Idef{org49}Institut f\"{u}r Informatik, Johann Wolfgang Goethe-Universit\"{a}t Frankfurt, Frankfurt, Germany
\item \Idef{org50}Institut f\"{u}r Kernphysik, Johann Wolfgang Goethe-Universit\"{a}t Frankfurt, Frankfurt, Germany
\item \Idef{org51}Institut f\"{u}r Kernphysik, Technische Universit\"{a}t Darmstadt, Darmstadt, Germany
\item \Idef{org52}Institut f\"{u}r Kernphysik, Westf\"{a}lische Wilhelms-Universit\"{a}t M\"{u}nster, M\"{u}nster, Germany
\item \Idef{org53}Institut Pluridisciplinaire Hubert Curien (IPHC), Universit\'{e} de Strasbourg, CNRS-IN2P3, Strasbourg, France
\item \Idef{org54}Institute for High Energy Physics, Protvino, Russia
\item \Idef{org55}Institute for Nuclear Research, Academy of Sciences, Moscow, Russia
\item \Idef{org56}Institute for Subatomic Physics of Utrecht University, Utrecht, Netherlands
\item \Idef{org57}Institute for Theoretical and Experimental Physics, Moscow, Russia
\item \Idef{org58}Institute of Experimental Physics, Slovak Academy of Sciences, Ko\v{s}ice, Slovakia
\item \Idef{org59}Institute of Physics, Academy of Sciences of the Czech Republic, Prague, Czech Republic
\item \Idef{org60}Institute of Physics, Bhubaneswar, India
\item \Idef{org61}Institute of Space Science (ISS), Bucharest, Romania
\item \Idef{org62}Instituto de Ciencias Nucleares, Universidad Nacional Aut\'{o}noma de M\'{e}xico, Mexico City, Mexico
\item \Idef{org63}Instituto de F\'{\i}sica, Universidad Nacional Aut\'{o}noma de M\'{e}xico, Mexico City, Mexico
\item \Idef{org64}iThemba LABS, National Research Foundation, Somerset West, South Africa
\item \Idef{org65}Joint Institute for Nuclear Research (JINR), Dubna, Russia
\item \Idef{org66}Konkuk University, Seoul, South Korea
\item \Idef{org67}Korea Institute of Science and Technology Information, Daejeon, South Korea
\item \Idef{org68}KTO Karatay University, Konya, Turkey
\item \Idef{org69}Laboratoire de Physique Corpusculaire (LPC), Clermont Universit\'{e}, Universit\'{e} Blaise Pascal, CNRS--IN2P3, Clermont-Ferrand, France
\item \Idef{org70}Laboratoire de Physique Subatomique et de Cosmologie (LPSC), Universit\'{e} Joseph Fourier, CNRS-IN2P3, Institut Polytechnique de Grenoble, Grenoble, France
\item \Idef{org71}Laboratori Nazionali di Frascati, INFN, Frascati, Italy
\item \Idef{org72}Laboratori Nazionali di Legnaro, INFN, Legnaro, Italy
\item \Idef{org73}Lawrence Berkeley National Laboratory, Berkeley, CA, United States
\item \Idef{org74}Lawrence Livermore National Laboratory, Livermore, CA, United States
\item \Idef{org75}Moscow Engineering Physics Institute, Moscow, Russia
\item \Idef{org76}National Centre for Nuclear Studies, Warsaw, Poland
\item \Idef{org77}National Institute for Physics and Nuclear Engineering, Bucharest, Romania
\item \Idef{org78}National Institute of Science Education and Research, Bhubaneswar, India
\item \Idef{org79}Niels Bohr Institute, University of Copenhagen, Copenhagen, Denmark
\item \Idef{org80}Nikhef, National Institute for Subatomic Physics, Amsterdam, Netherlands
\item \Idef{org81}Nuclear Physics Group, STFC Daresbury Laboratory, Daresbury, United Kingdom
\item \Idef{org82}Nuclear Physics Institute, Academy of Sciences of the Czech Republic, \v{R}e\v{z} u Prahy, Czech Republic
\item \Idef{org83}Oak Ridge National Laboratory, Oak Ridge, TN, United States
\item \Idef{org84}Petersburg Nuclear Physics Institute, Gatchina, Russia
\item \Idef{org85}Physics Department, Creighton University, Omaha, NE, United States
\item \Idef{org86}Physics Department, Panjab University, Chandigarh, India
\item \Idef{org87}Physics Department, University of Athens, Athens, Greece
\item \Idef{org88}Physics Department, University of Cape Town, Cape Town, South Africa
\item \Idef{org89}Physics Department, University of Jammu, Jammu, India
\item \Idef{org90}Physics Department, University of Rajasthan, Jaipur, India
\item \Idef{org91}Physikalisches Institut, Ruprecht-Karls-Universit\"{a}t Heidelberg, Heidelberg, Germany
\item \Idef{org92}Politecnico di Torino, Turin, Italy
\item \Idef{org93}Purdue University, West Lafayette, IN, United States
\item \Idef{org94}Pusan National University, Pusan, South Korea
\item \Idef{org95}Research Division and ExtreMe Matter Institute EMMI, GSI Helmholtzzentrum f\"ur Schwerionenforschung, Darmstadt, Germany
\item \Idef{org96}Rudjer Bo\v{s}kovi\'{c} Institute, Zagreb, Croatia
\item \Idef{org97}Russian Federal Nuclear Center (VNIIEF), Sarov, Russia
\item \Idef{org98}Russian Research Centre Kurchatov Institute, Moscow, Russia
\item \Idef{org99}Saha Institute of Nuclear Physics, Kolkata, India
\item \Idef{org100}School of Physics and Astronomy, University of Birmingham, Birmingham, United Kingdom
\item \Idef{org101}Secci\'{o}n F\'{\i}sica, Departamento de Ciencias, Pontificia Universidad Cat\'{o}lica del Per\'{u}, Lima, Peru
\item \Idef{org102}Sezione INFN, Bari, Italy
\item \Idef{org103}Sezione INFN, Bologna, Italy
\item \Idef{org104}Sezione INFN, Cagliari, Italy
\item \Idef{org105}Sezione INFN, Catania, Italy
\item \Idef{org106}Sezione INFN, Padova, Italy
\item \Idef{org107}Sezione INFN, Rome, Italy
\item \Idef{org108}Sezione INFN, Trieste, Italy
\item \Idef{org109}Sezione INFN, Turin, Italy
\item \Idef{org110}SUBATECH, Ecole des Mines de Nantes, Universit\'{e} de Nantes, CNRS-IN2P3, Nantes, France
\item \Idef{org111}Suranaree University of Technology, Nakhon Ratchasima, Thailand
\item \Idef{org112}Technical University of Split FESB, Split, Croatia
\item \Idef{org113}Technische Universit\"{a}t M\"{u}nchen, Munich, Germany
\item \Idef{org114}The Henryk Niewodniczanski Institute of Nuclear Physics, Polish Academy of Sciences, Cracow, Poland
\item \Idef{org115}The University of Texas at Austin, Physics Department, Austin, TX, USA
\item \Idef{org116}Universidad Aut\'{o}noma de Sinaloa, Culiac\'{a}n, Mexico
\item \Idef{org117}Universidade de S\~{a}o Paulo (USP), S\~{a}o Paulo, Brazil
\item \Idef{org118}Universidade Estadual de Campinas (UNICAMP), Campinas, Brazil
\item \Idef{org119}University of Houston, Houston, TX, United States
\item \Idef{org120}University of Jyv\"{a}skyl\"{a}, Jyv\"{a}skyl\"{a}, Finland
\item \Idef{org121}University of Liverpool, Liverpool, United Kingdom
\item \Idef{org122}University of Tennessee, Knoxville, TN, United States
\item \Idef{org123}University of Tokyo, Tokyo, Japan
\item \Idef{org124}University of Tsukuba, Tsukuba, Japan
\item \Idef{org125}University of Zagreb, Zagreb, Croatia
\item \Idef{org126}Universit\'{e} de Lyon, Universit\'{e} Lyon 1, CNRS/IN2P3, IPN-Lyon, Villeurbanne, France
\item \Idef{org127}V.~Fock Institute for Physics, St. Petersburg State University, St. Petersburg, Russia
\item \Idef{org128}Variable Energy Cyclotron Centre, Kolkata, India
\item \Idef{org129}Vestfold University College, Tonsberg, Norway
\item \Idef{org130}Warsaw University of Technology, Warsaw, Poland
\item \Idef{org131}Wayne State University, Detroit, MI, United States
\item \Idef{org132}Wigner Research Centre for Physics, Hungarian Academy of Sciences, Budapest, Hungary
\item \Idef{org133}Yale University, New Haven, CT, United States
\item \Idef{org134}Yonsei University, Seoul, South Korea
\item \Idef{org135}Zentrum f\"{u}r Technologietransfer und Telekommunikation (ZTT), Fachhochschule Worms, Worms, Germany
\end{Authlist}
\endgroup

\end{document}